\documentclass[a4paper]{article}

\usepackage
[
a4paper,
left=1in,
right=1in,
top=1in,
bottom=1in,
]
{geometry}
\usepackage{lipsum}
\usepackage{graphicx}
\usepackage{amsmath}
\usepackage{amssymb}
\usepackage{soul}
\usepackage[center]{caption}
\usepackage[export]{adjustbox}
\usepackage{array}
\usepackage[caption=false,font=footnotesize]{subfig}

\newenvironment{blockquote}{%
	\par%
	\leftskip=1em
	\noindent\ignorespaces}{%
	\par}

\providecommand{\keywords}[1]
{
	\small	
	\textbf{\textit{Keywords---}} #1
}

\title{Energy-Optimal Control of a Submarine-Launched Cruise Missile\thanks{This work has been submitted to the IEEE for possible publication. Copyright may be transferred without notice, after which this version may no longer be accessible.\newline \hspace*{0.18in}E-mail addresses: semih.koklucan@metu.edu.tr (corresponding author), kleb@metu.edu.tr}}
\author{Semih K\=okl\=ucan$^{1, 2}$, M. Kemal Leblebicio\u{g}lu$^{1}$  \\ \\
	\footnotesize $^{1}$Department of Electrical and Electronics Engineering, Middle East Technical University, Ankara, 06800, Turkey \\
	\footnotesize $^{2}$The Scientific and Technological Research Council of Turkey-Defense Industries Research and Development \\ \footnotesize  Institute (T\"UB\.ITAK-SAGE), Ankara, 06261, Turkey \\
}
\date{}
\begin{document}

\maketitle

\begin{abstract}
A typical mission profile of submarine-launched cruise missiles begins with the launch
phase which covers the motion of the missile from the launch to the water-exit and
continues with the boost phase which lasts from the water-exit to the beginning of the
cruise phase. In order to achieve the desired range of the launch and boost phases,
efficient utilization of available energy which carries the missile to the beginning of the cruise phase is necessary. For this purpose, this study presents a new approach
for energy-optimal control of the underwater and air motion of a submarine-launched
cruise missile. In this approach, the aforementioned problem is modeled and solved
as a minimum-effort optimal control problem. Then, the effects of initial and final
conditions on energy need are investigated, and the optimal conditions that result
with the minimum energy need are determined. Prior to the guidance and control design steps, six degrees of freedom (6 DOF) motion equations are derived and the hydrodynamic and aerodynamic parameters are retrieved. The nonlinear 6 DOF motion model is simplified and linearized before minimum-effort optimal control design part. Results of the designed guidance and control strategies are presented through the nonlinear 6 DOF simulations. Finally, some comments are made and future studies are mentioned based on theoretical and simulation studies.\\\\

\end{abstract}

\section{Introduction}
%
%
%
%

Cruise missiles are precision-guided weapons which are able to fly long distances to accomplish strategic or tactical missions. Through their flight, they are usually continuously powered by an engine and supported by aerodynamic surfaces to make use of the aerodynamic lift. Aircraft, land-based launchers, ships or submarines can be used as launching platforms and cruise missiles can be used against land or sea targets. From the beginning of their first development in the 1940s, today's modern cruise missiles are equipped with advanced technologies, and they can play a crucial role in defense organizations of countries. Underwater launching of cruise missiles differs from other launching methods in several important ways. Since the electromagnetic wave propagation is strongly attenuated by the seawater, radars cannot detect the submarines efficiently. Furthermore, the available sonar systems of today are not able to provide efficient detection ranges for modern submarine systems. The underwater operation can hide submarines from other reconnaissance missions performed by satellites and aerial vehicles. Therefore, concealment, sudden attack ability and high level of survivability are some aspects of submarine-launched missiles \cite{zhang2018fixed}, \cite{sonarsystems}.

Typically, two phases of a submarine-launched cruise missile are seen before the cruise phase begins. These phases are the launch phase, which covers the motion in the underwater, and the boost phase, which covers the motion in the air until the beginning of the cruise phase. During the launch and boost phases, the commonly used control method is generating necessary forces and moments via thrust control. In general, these can be achieved by vectoring booster motor thrust or using additional lateral thrusters. Aerodynamic surfaces, which are less effective than using thrust forces and moments where aerodynamic force is low, may also be used in combination with thrust control  
\cite{hong2019nonlinear}--\cite{wassom1991}.

In order to achieve the desired range or increase the range of the launch and boost phases, it is crucial to use available energy for thrust generation efficiently. The total energy need of the launch and boost phases can change according to the initial and final flight conditions of these phases and also with the total planned completion time. Thus, a control and guidance scheme to be applied should take into account these factors to increase the missile range for the phases before starting to cruise. 

Some notable works related to the guidance and control of underwater-launched missiles can be listed as follows. In \cite{haochun2013guidance}, attitude and depth/height control of a  submarine-to-air missile are achieved using Linear Quadratic  Regulator and an optimal guidance law.  Time-optimal control of a submarine-launched missile based on thrust vector deflection parameterization with the enhanced time-scaling method is given in \cite{wen2016time}. Sliding mode based water-exit attitude controllers for submarine-launched missiles are presented in \cite{zhang2018fixed} and \cite{crossmedium2018}. In addition to these works, there are some studies which focus on water-exit or just after water-exit control of some vehicles operationally and physically similar to missiles. Reference \cite{xiao2013modeling} and \cite{minxiao2} are studies in which a nonlinear sliding mode controller utilizing an adaptive backstepping approach and a variable structure controller are designed in order to achieve attitude tracking control for out-of-water course of high-speed underwater vehicles, respectively. Reference \cite{qi2017water} suggests a sliding mode controller which guarantees that an underwater-to-air vehicle reaches the desired conditions after leaving the water and it prevents the vehicle from falling back into the water. 

Moreover, there are also some important studies that do not include a controller design but concentrate on modeling and simulation of underwater launch and water-exit phases of missiles. The water-exit dynamics are modeled and motion simulations are carried out in \cite{yang2017water}--\cite{wu2019}. References \cite{shang2012}--\cite{weiland2010} present modeling and simulation studies for underwater launch phase for various conditions. Reference \cite{peng2018} is another study that focused on operation concept time optimization of submarine-launched missiles which attack sea targets. Besides them, there are some important works which focus on modeling and simulation studies for water-exit motion of some vehicles operationally and physically similar to missiles. In these works, the dynamic model for water-exit of slender vehicles \cite{anliu2020}, \cite{fu2018}, a morphing unmanned submersible aerial vehicle \cite{junhua2017}, a cylinder vehicle \cite{bxu2015}, submarine-launched vehicles \cite{wang2018}, \cite{zahid2020}, are developed and motion simulations are performed.

One aspect that is not dealt with in the aforementioned works is energy optimizing control in which the energy need for the thrust generation is considered. As it is previously explained, this can be a desirable objective in order to achieve the desired range or increase the range for the launch and boost phases. Obtaining some predefined flight conditions at the end of the boost phase may be necessary for real systems due to mission objectives or system constraints. For example, the cruise engine can only start within a specific altitude-speed envelope, or initial success of the cruise autopilot may depend on the flight conditions at which the boost phase ends. Therefore, effective utilization of the available energy to increase missile range may give flexibility to the system designer to accomplish some other goals. 

In our new approach, to suggest a solution to the energy optimization problem, launch and boost phase control problems of a conceptual submarine-launched cruise missile are modeled and solved as a minimum-effort optimal control problem. Formulated optimal control problems are transformed into optimization problems by discretizing them and then solved as parameter optimization problems. The energy associative cost function to be minimized is determined as the integral of the square of the applied thrust. Firstly, the launch and boost phases are considered separately, and energy minimizing control solutions are found to satisfy the given initial and final conditions. These conditions include velocity, depth/altitude, attitude, and time interval. Secondly, the effects of the final and initial conditions on total energy need are investigated. The optimal conditions that result with the minimum energy need are determined. Where it is appropriate, initial and final conditions are also used as free parameters to be found, to minimize the control effort. Then, for the eventual initial and final conditions of the launch and boost phases, energy minimizing control solution and resulting flight scenario is obtained for these phases. 

Some limited parts of the works presented in this article are previously published in two national conferences in Turkish \cite{usmos2019}, \cite{tok2019}. However, these publications only focused on finding optimal solutions for launch phase different-water exit pitch angle scenarios of horizontal and horizontal/vertical launches, respectively. The works in this article are substantially extended by finding both launch and boost phase solutions together, investigating the effects of the final and initial conditions on total energy need, and obtaining the optimal initial and final conditions for overall design. 

The remaining parts of this paper are organized as follows. In Section \ref{sec:s2}, the conceptual missile system design is described with its flight modes, control methods, and physical properties. Section \ref{sec:s3} is dedicated to the mathematical modeling of the motion of the missile in 6 degrees of freedom and the retrieval of related model parameters. In Section \ref{sec:s4}, minimum-effort control design for launch and boost phases are accomplished. Lastly, the results of the study are discussed and conclusions are given in Section \ref{sec:s5}.


\section {Missile System Description}
\label{sec:s2}
A conceptual cruise missile system is designed in this work. There are two different flight modes of the missile which accompany the mission phases that will be studied here. The first flight mode is active through the launch phase, which covers the motion of the missile from submarine ejection to the water-exit. The second flight mode is active through the boost phase, which covers the motion from the sea surface to the beginning of the cruise phase.

\begin{figure}[!t]
	\centering
	\includegraphics[width=2.5in]{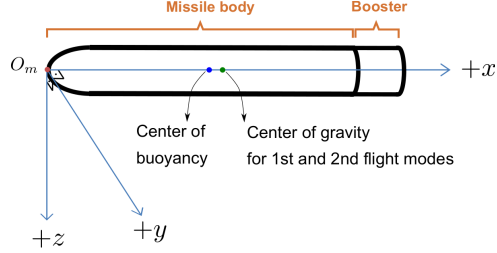}
	\caption{2-dimensional view of the conceptual missile design}
	\label{fig:cg_cb_locations}
\end{figure}

The 2D view of the missile with its center of gravity and the center of buoyancy locations on a measurement frame centered at $O_m$ are given in Fig. \ref{fig:cg_cb_locations}. Missile main body and the booster motor has a cylindrical shape, whereas the nose is in the form of ellipsoid.

\begin{figure}[!t]
	\centering
	\includegraphics[width=3.0in]{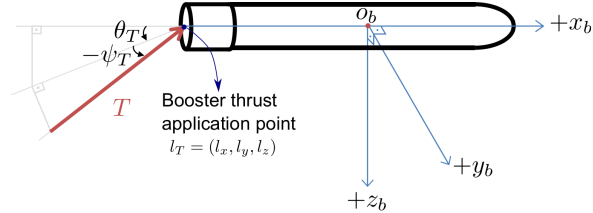}
	\caption{Application point of booster thrust and thrust deflection angles}
	\label{fig:tvc_app_point}
\end{figure}

In this study, the missile will be used for both the horizontal and the vertical launch scenarios. Following the horizontal ejection from a submarine, the missile has a tendency to pitch upwards in the underwater. This is possible with the position relationship between the center of buoyancy and the center of gravity. Since the center of buoyancy is closer to the nose of the missile than that of the center of gravity, buoyancy force generates an upwards pitching moment for the nose. By making use of this design characteristic, water-exit attitude control will be achieved by igniting the missile booster at a predetermined condition while the missile is pitching upwards, and then applying a predetermined booster thrust profile. By doing so, the control will be achieved without any thrust deflection in the underwater. For the vertical ejection scenarios, vertical water-exit will be desired. During the boost phase, thrust vector control (TVC) will be used to control the attitude of the missile. The point of application of the booster thrust on the missile body frame, which is centered at $O_b$, and possible thrust deflections are shown in Fig. \ref{fig:tvc_app_point}. 

In practice, tail fins could be necessary during launch and boost phases for purposes such as roll control or additional control elements for attitude control. However, in the scope of this work, only longitudinal motion control will be considered and it will be assumed that there is not any motion in roll and yaw dynamics. So, tail fin control is not included in the design. To form a basis for the physical parameters of the missile, the Tomahawk cruise missile is chosen. For the data which is not available in the related open sources, reasonable values are chosen or calculated according to the physical model of the missile. Related physical parameters for two different flight modes are given in Table \ref{table_phys_params}. The center of gravity, $(x_{cg}, y_{cg}, z_{cg})$, and the center of buoyancy, $(x_{cb}, y_{cb}, z_{cb})$, locations are given in the measurement frame which is shown in Fig. \ref{fig:cg_cb_locations}. $(I_x, I_y, I_z)$ are the moments of inertia about the $x$, $y$, and $z$ axes of the missile body frame. All parameter values are assumed to be constant and same for the first and second flight modes by ignoring some realistic facts such as the decrease of the total mass due to the booster (fuel consumption) or any physical change during the transition between the flight modes, such as separation of some physical parts after the water-exit.

\begin{table}[!t]
	\renewcommand{\arraystretch}{1.3}
	\caption{Physical Parameters for the Conceptual Missile Design}
	\label{table_phys_params}
	\centering
	\begin{tabular}{|c|c|c|}
		\hline
		\textbf{Parameter} & \textbf{Value} & \textbf{Unit}\\
		\hline
		Length & 6.1806 & $m$ \\
		Diameter & 0.5175 &  $m$ \\
		Mass & 1513 & $kg$ \\ 
		Reference Area & 0.2104 & $m^2$ \\
		Volume & 1.332 & $m^3$ \\ 
		$(I_x, I_y, I_z)$ & (50.6684, 4841.6944, 4841.6944) & $kgm^2$ \\ 
		$(x_{cg}, y_{cg}, z_{cg})$ & (3.1903, 0, 0) & $m$ \\
		$(x_{cb}, y_{cb}, z_{cb})$ & (3.0903, 0, 0) & $m$ \\ 
		\hline
	\end{tabular}
\end{table}


\section {Mathematical Modeling}
\label{sec:s3}

\subsection{Six Degrees of Freedom (6 DOF) Equations of Motion}
\begin{figure}[!t]
	\centering
	\includegraphics[width=2.0in]{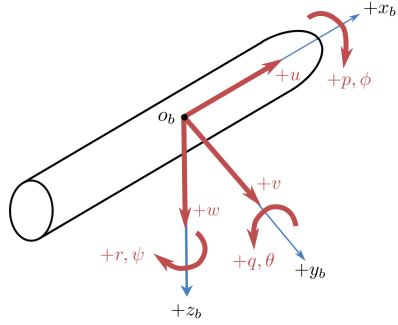}
	\caption{The 6 DOF motion variables in the body frame}
	\label{fig:6dof_motion_vars}
\end{figure}

The body frame of the missile and some motion variables related to the 6 DOF motion model are shown in Fig. \ref{fig:6dof_motion_vars}. Here $(u, v, w)$ are the forward, side, and down velocities in the body frame, $(\phi, \theta, \psi)$ are the euler angles which describe roll, pitch, and yaw angles with respect to a reference frame, $(p, q, r)$ are the angular velocities about the $x$, $y$, and $z$ axes of the body frame. Due to the $xy$-plane and $xz$-plane symmetries of the missile in the body frame, and by choosing the body frame origin as the center of gravity, dynamical equations of motion can be stated as \cite{fossen1994guidance}
\begin{equation}
\label{eqn:eom_first}
\begin{split}
m(\dot{u}-vr+wq)=X \quad &I_x\dot{p}+(I_z-I_y)qr=L  \\
m(\dot{v}-wp+ur)=Y \quad &I_y\dot{q}+(I_x-I_z)pr=M  \\
m(\dot{w}-uq+vp)=Z \quad &I_z\dot{r}+(I_y-I_x)pq=N
\end{split}
\end{equation}
where $(X,Y,Z)$ and $(L,M,N)$ are total forces and moments applied to the missile in the body frame, $m$ is mass, and $(I_x,I_y,I_z)$ are the moments of inertia. If the total forces and moments, and the velocity variables are defined in vector form as
\begin{equation}
\label{eqn:force_velocity_vector}
\boldsymbol{\tau} = [X,Y,Z,L,M,N]^T \quad \boldsymbol{\nu} = [u,v,w,p,q,r]^T
\end{equation}
then, the compact vector form of \eqref{eqn:eom_first} is expressed as
\begin{equation}
\label{eqn:eom_second}
\boldsymbol{M_{RB}}\dot{\boldsymbol{\nu}} + \boldsymbol{C_{RB}(\nu)\nu}= \boldsymbol{\tau}
\end{equation}
where $\boldsymbol{M_{RB}}$ is rigid body mass matrix, and $\boldsymbol{C_{RB}(\nu)}$ is rigid body Coriolis and centripetal matrix.

\subsection{Forces and Moments}
The total forces and moments which act on the missile body can be considered as the sum of the following factors: aerodynamic and hydrodynamic forces and moments including added mass and inertia effect, gravitational and buoyancy forces and moments, and thrust forces and moments \cite{fossen1994guidance}. Environmental sources such as wind, waves, and currents are assumed to be zero within the scope of this work. Morevor, motion model is derived by assuming that the missile is either completely in underwater or air. So, the water-exit dynamics that can be seen during the transition from the underwater to air is ignored.

\subsubsection{Aerodynamic and Hydrodynamic Forces and Moments}
\underline{Added Mass and Inertia:} The motion of a rigid body in a fluid causes the fluid surrounding the body is accelerated together with the body. As a result of this, a force is needed to achieve this acceleration with which the fluid reacts with a force that is equal in magnitude and opposite in direction. This reaction force is named as added mass contribution. For example, the hydrodynamic added mass force along $x_b$-axis as a result of linear acceleration $\dot u$ in the $x_b$ direction is written as \cite{antonelli2014underwater}
\begin{equation}
X_A = - X_ {\dot u} \dot u, \hspace{1cm} where \hspace{0.5cm}  X_ {\dot u} = \frac{\partial X}{\partial \dot u}
\end{equation}
Considering the $xy$-plane and $xz$-plane symmetries of the missile, added mass effects can be included in the motion equations with the matrices given in \eqref{eqn:added_mass_matrices} \cite{rentschler2003dynamic}
\begin{equation}
\label{eqn:added_mass_matrices}
\begin{split}
\boldsymbol M_A & \triangleq -
\left[
\begin{matrix}
X_{\dot u} & 0 & 0 & 0 & 0 & 0 \\
0 & Y_{\dot v} & 0 & 0 & 0 & Y_{\dot r} \\
0 & 0 & Z_{\dot w} & 0 & Z_{\dot q} & 0 \\
0 & 0 & 0 & K_{\dot p} & 0 & 0 \\
0 & 0 & M_{\dot w} & 0 & M_{\dot q} & 0 \\
0 & N_{\dot v} & 0 & 0 & 0 & N_{\dot r} \\
\end{matrix}
\right]\\
\boldsymbol C_A(\boldsymbol \nu) &= 
\left[
\begin{matrix}
0 & 0 & 0 & 0 & -a_3 & a_2 \\
0 & 0 & 0 & a_3 & 0 & -a_1 \\
0 & 0 & 0 & -a_2 & a_1 & 0 \\
0 & -a_3 & a_2 & 0 & -b_3 & b_2 \\
a_3 & 0 & -a_1 & b_3 & 0 & -b_1 \\
-a_2 & a_1 & 0 & -b_2 & b_1 & 0 \\
\end{matrix}
\right]
\end{split}
\end{equation}
where $\boldsymbol{M_{A}}$ and $\boldsymbol{C_{A}(\nu)}$ are the mass and Coriolis and centripetal terms related matrices that will be added in \eqref{eqn:eom_second}. Coriolis and centripetal matrix elements can be written as
\begin{equation}
\begin{split}
a_1 = X_{\dot u}u \quad & b_1  = K_{\dot p}p \\
a_2 = Y_{\dot v}v + Y_{\dot r}r \quad & b_2  = M_{\dot w}w + M_{\dot q}q \\
a_3 = Z_{\dot w}w + Z_{\dot q}q \quad & b_3  = N_{\dot v}v + N_{\dot r}r \\
\end{split}
\end{equation}
\ul{Representation of Aerodynamic and Hydrodynamic Forces and Moments:}
The contributions of aerodynamic and hydrodynamic forces and moments can be represented in vectorial form as
\begin{equation}
\boldsymbol{\tau_{A/H}} = QA[C_x \ C_y \ C_z \ dC_l \ dC_m \ dC_n]^T
\end{equation}
where $C_x, C_y, C_z$ are force coefficients, $C_l, C_m, C_n$ are moment coefficients, $Q$ is dynamic pressure, $A$ is reference area, and $d$ is reference length. Dynamic pressure is defined as:
\begin{equation}
Q = \frac{1}{2}\rho V^2
\end{equation}
where $V$ is the magnitude of the total velocity of the missile and can be found by the formula:
\begin{equation}
V = \sqrt {u^2 + v^2 + w^2}
\end{equation}
Depending on the missile's surrounding medium, $\rho$ is the water or air density. In this work, water density is taken as the sea water density, 1023 $kg/m^3$, and the air density is calculated according to the International Standard Atmosphere model.

In the scope of this work, force and moment coefficients are represented in a linear form as
\begin{equation}
\begin{split}
C_i = C_{i0} + C_{ip}p(d/2V) + C_{iq}q(d/2V) + C_{ir}r(d/2V) \\
for \quad i=x,y,z,l,m,n
\end{split}
\end{equation}
where $C_{i0}$ are static coefficients, and $C_{ip}$, $C_{iq}$, $C_{ir}$ are dynamic moment coefficients.

\subsubsection{Gravitational and Buoyancy Forces and Moments}
The gravitational and buoyancy forces are called restoring forces in hydrostatic terminology. Through this study $\boldsymbol \tau_R$ will be used to denote restoring forces. Assuming the center of gravity and the center of buoyancy are on the $x$ axis of the body frame, and the body frame origin is the center of gravity, total restoring forces and moments can be written as
\begin{equation}
\boldsymbol \tau_R=  
\left[
\begin{matrix}
(B - W)\sin(\theta) \\
(W - B)\cos(\theta)\sin(\phi) \\
(W - B)\cos(\theta)\cos(\phi) \\
0 \\
x_bB\cos(\theta)\cos(\phi) \\
-x_bB\cos(\theta)\sin(\phi)\\
\end{matrix} 
\right]
\end{equation}
where $x_b$ is the distance between the center of buoyancy and the center of gravity, $W$ is weight of the missile, and $B$ is buoyancy force that is exerted to the missile by the water.

\subsubsection{Thrust Forces and Moments}

The thrust forces and moments which act on the missile body are produced by a booster motor. Thrust force and moment vector $\boldsymbol \tau_T$ can be expressed as
\begin{equation}
\label{eqn_thrust_f_m_1}
\boldsymbol \tau_T = 
\left[
\begin{matrix}
T\cos(\theta_T)\cos(\psi_T) \\
T\sin(\psi_T) \\
-T\sin(\theta_T)\cos(\psi_T) \\
-Tl_y\sin(\theta_T)\cos(\psi_T)-Tl_z\sin(\psi_T) \\
Tl_z\cos(\theta_T)\cos(\psi_T)+Tl_x\sin(\theta_T)\cos(\psi_T) \\
-Tl_y\cos(\theta_T)\cos(\psi_T)+Tl_x\sin(\psi_T)
\end{matrix}
\right]
\end{equation}
Here, $T$ is total thrust, $\theta_T$ is thrust deflection in pitch plane, $\psi_T$ is thrust deflection in yaw plane, and $l_x, l_y, l_z$ are the components of $\boldsymbol l_T$, which is the moment arm of booster thrust vector. If it is assumed that thrust deflections are small and thrust is applied at a point of $x$ axis of the body frame, then $\boldsymbol l_T = [l_x, 0, 0]$, and \eqref{eqn_thrust_f_m_1} can be simplified as \begin{equation}
\boldsymbol \tau_T = 
\left[
\begin{matrix}
T \ T\psi_T \ -T\theta_T \ 0 \ Tl_x\theta_T \ Tl_x\psi_T
\end{matrix}
\right]^T
\end{equation}

\subsubsection{Conclusion}

As a result of the force and moment sources that are described in the previous sections, the equations of motion given in \eqref{eqn:eom_second} can be rewritten as
\begin{equation}
\label{eom_final}
(\boldsymbol{M_{RB}}+\boldsymbol{M_{A}})\dot{\boldsymbol{\nu}} + (\boldsymbol{C_{RB}(\nu)}+\boldsymbol{C_{A}(\nu)})\nu= \boldsymbol{\tau_{A/H}} + \boldsymbol{\tau_R} + \boldsymbol{\tau_T}
\end{equation}
where the explicit definitions of each term were given previously in the corresponding sections.

\subsection{Derivation of Hydroynamic and Aerodynamic Parameters}

Having expressed the aerodynamic and hydrodynamic forces and moments, numerical values of hydrodynamic and aerodynamic parameters that are involved in these expressions should be derived. For this purpose, added mass parameters and hydrodynamic/aerodynamic force and moment coefficients are calculated.

\subsubsection{Added Mass Parameters}
They are the parameters which are included in the added mass matrix and Coriolis and centripetal matrix related to added mass effect, given in \eqref{eqn:added_mass_matrices}. Due to the simplifications of these matrices because of the physical properties of the missile, the parameters to be found reduce to $X_{\dot{u}}$, $Y_{\dot{v}}$, $Y_{\dot{r}}$, $Z_{\dot{w}}$, $Z_{\dot{q}}$, $K_{\dot{p}}$, $M_{\dot{w}}$, $M_{\dot{q}}$, $N_{\dot{v}}$, $N_{\dot{r}}$. In order to obtain their numerical values, the methods given in \cite{rentschler2003dynamic} is used. Axial added mass parameter, $X_{\dot{u}}$, is found with the approach of Blevins by approximating the missile shape as an ellipsoid \cite{blevins1980formulas}. Rolling added mass, $K_{\dot{p}}$, is assumed to be zero since there is no fin or another control surface on the cylindrical shape of the missile. Remaining added mass parameters can be found using the strip theory \cite{fossen2011handbook}, \cite{newman2018marine}. As the result, the calculated added mass parameters are given in Table \ref{table:added_mass_params}.

\begin{table}[!t]
	\renewcommand{\arraystretch}{1.3}
	\caption{Added Mass Parameters}
	\label{table:added_mass_params}
	\centering
	\resizebox{.45\textwidth}{!}{%
		\begin{tabular}{|c|c|c||c|c|c|}
			\hline
			\textbf{Parameter} & \textbf{Value} & \textbf{Unit} & \textbf{Parameter} & \textbf{Value} & \textbf{Unit} \\
			\hline
			$X_{\dot{u}}$ & -10.5294 	& $kg$ & $K_{\dot{p}}$ & 0 & $\frac{kgm^2}{rad}$ \\
			$Y_{\dot{v}}$ & -1296.5 	& $kg$ & $M_{\dot{w}}$ &  99.4382 	& $kgm$ \\
			$Y_{\dot{r}}$ & -99.4382 	& $\frac{kgm}{rad}$ & $M_{\dot{q}}$ & -3936.7  	& $\frac{kgm^2}{rad}$ \\ 
			$Z_{\dot{w}}$ & -1296.5 	& $kg$ & $N_{\dot{v}}$ & -99.4382 	& $kgm$ \\ 
			$Z_{\dot{q}}$ &  99.4382 	& $\frac{kgm}{rad}$ & $N_{\dot{r}}$ & -3936.7 	& $\frac{kgm^2}{rad}$ \\ 							
			\hline
		\end{tabular}
	}
\end{table}

\subsubsection{Hydrodynamic and Aerodynamic Force and Moment Coefficients}
The numerical values of the hydrodynamic and aerodynamic force and moment coefficients, $(C_{i0}, C_{ip}, C_{iq}, C_{ir}\ |\ i=x,y,z,l,m,n)$, are obtained by using Missile DATCOM software \cite{rosema2011missile}. The coefficients are obtained in a tabulated form where the coefficient values are listed according to the angle of attack, sideslip angle and Mach number values of the missile at a specific flight condition. During the simulations, each coefficient is calculated at the corresponding flight condition by interpolation.

\section {Minimum-Effort Optimal Control Design}
\label{sec:s4}
In this study, the energy optimizing guidance and control problem of the launch and boost phases have been dealt with as a minimum-effort optimal control problem. First, simplified equations of motion are obtained for both launch and boost phases. Then, the minimum-effort optimal control problem is formulated mathematically. In that formulation, energy associative cost function to be minimized is determined as the integral of the square of the applied thrust. To be able to solve the formulated infinite dimensional optimal control problems numerically, they are treated as finite dimensional parameter optimization problems. At first, optimal control solutions for different water-exit pitch angle scenarios are obtained. After that, the effect of other initial and final conditions on the cost are investigated and optimal conditions that result with minimum energy scenarios are found. Finally, the cost analysis for overall design is stated, and the mission profiles for horizontal and vertical launch cases with minimum energy are described.

\subsection{Simplified Equations of Motion}
Simplification of the nonlinear equations of motion for launch and boost phases is
aimed for several reasons. Firstly, in the scope of this work, for both of the phases, no control is desired in the lateral plane. Assuming that there is not any lateral motion or a disturbance, longitudinal dynamics can be decoupled from lateral dynamics. Then, it will be sufficient only to consider the control effort for longitudinal dynamics. Secondly, the optimal control solutions will be found using optimization, but the performance of the optimization algorithm decreases as the model complexity increases. To decrease the computation time of optimization algorithms, it is advantageous to have simpler models. Therefore, nonlinear equations of motion for the launch and boost phases are simplified.

\subsubsection{Simplified Equations of Motion for Launch Phase}
In this phase, control action only consists of changing booster thrust magnitude. As it is previously explained, due to the relationship between the center of buoyancy and the center of gravity, when the missile is at rest horizontally in underwater, a positive pitching moment occurs. Then, if an appropriate thrust profile is applied after the horizontal ejection, while the missile is being carried up to the sea surface, attitude and water-exit velocity can also be controlled. In this phase, no control is desired in the lateral plane. By assuming no disturbance for lateral motion, longitudinal motion can be decoupled from lateral motion. In summary, the following assumptions can be made to simplify the dynamics of the launch phase:

\begin{itemize}
	\item Thrust deflection angles, $\theta_{T}$ and $\psi_{T}$, are zero.
	\item Roll rate, $p$, and yaw rate, $r$, are zero.
	\item Side body velocity, $v$, is zero.
	\item Among body velocities, forward velocity is the dominant component, i.e., $u >> v$ and $u >> w$.
\end{itemize}
Hence, using \eqref{eom_final}, equations of motion related to longitudinal dynamics become
\begin{equation}
\label{eqn:chapter5_launch_pitch_eqns}
\begin{split}
X &= m\dot{u} - X_{\dot{u}}\dot{u} + mwq - Z_{\dot{w}}qw - Z_{\dot{q}}q^2 \\
Z &= m\dot{w} - Z_{\dot{w}}\dot{w} - Z_{\dot{q}}\dot{q} - mqu + X_{\dot{u}}uq\\
M &= I_y\dot{q} - M_{\dot{w}}\dot{w} - M_{\dot{q}}\dot{q} + Z_{\dot{w}}wu + Z_{\dot{q}}qu - X_{\dot{u}}uw
\end{split}
\end{equation}
Moreover, with the assumptions above, force and moment terms are simplified as
\begin{equation}
\label{eqn:chapter5_launch_force_eqns}
\begin{split}
X &= QA(C_{x0} + C_{xq}\frac{d}{2u}q)+T-(W-B)\sin(\theta)\\
Z &= QA(C_{z0} + C_{zq}\frac{d}{2u}q)+(W-B)\cos(\theta)\\
M &= QAd(C_{m0} + C_{mq}\frac{d}{2u}q)+x_bB\cos(\theta)\\
\end{split}
\end{equation}
Substituting \eqref{eqn:chapter5_launch_force_eqns} into \eqref{eqn:chapter5_launch_pitch_eqns} and considering the additional states related to depth, $z$, and pitch angle, $\theta$, launch phase dynamics can be obtained as
\begin{equation}
\label{eqn:chapter5_launch_state_eqns_1}
\begin{split}
\dot{u} &= (QAC_{x0} + QAC_{xq}(d/2u)q+T-(W-B)\sin\theta \\
& \quad - mwq +Z_{\dot{w}}wq+Z_{\dot{q}}q^2)/(m-X_{\dot{u}})\\
\dot{w} &= (QAC_{z0}+QAC_{zq}(d/2u)q+(W-B)\cos\theta \\ 
& \quad + mqu + Z_{\dot{q}}\dot{q}-X_{\dot{u}}qu)/(m-Z_{\dot{w}}) \\
\dot{q} &= (QAdC_{m0}+QAC_{mq}(d^2/2u)q+x_bB\cos\theta+M_{\dot{w}}\dot{w}\\
&\quad  -Z_{\dot{w}}wu+Z_{\dot{q}}qu+X_{\dot{u}}uw)/(I_y-M_{\dot{q}})\\
\dot{\theta} &= q\\
\dot{z} &= -\sin(\theta)u+\cos(\theta)w
\end{split}
\end{equation}
In conclusion, longitudinal dynamics of the missile in launch phase can be expressed as
\begin{equation}
\label{eqn:chapter5_launch_state_eqns_final}
\boldsymbol{\dot{x}} = \boldsymbol{h} (\boldsymbol{x},\boldsymbol{u})
\end{equation}
where the state vector $\boldsymbol{x}$ and control vector $\boldsymbol{u}$ are,
\begin{equation}
\begin{split}
\boldsymbol{x} &= [u(t),w(t),q(t),\theta(t),z(t)]^T \\
\boldsymbol{u} &= [T(t)]
\end{split}
\end{equation}
and $\boldsymbol{h} (\boldsymbol{x},\boldsymbol{u})$ is the vector of expressions given in right-hand side of \eqref{eqn:chapter5_launch_state_eqns_1}.

\subsubsection{Simplified Equations of Motion for Boost Phase}
In this phase, control action consists of changing both booster thrust magnitude and thrust deflection. Attitude control of this phase will be achieved by using thrust vector control. As in the case of the launch phase, no control is desired in the lateral plane for the boost phase. By assuming no disturbance for lateral motion, longitudinal motion can be decoupled from lateral motion. In summary, the following assumptions can be made to simplify the dynamics of the launch phase:
\begin{itemize}
	\item Roll rate, $p$, and yaw rate, $r$, are zero.
	\item Side body velocity, $v$, is zero.
	\item Among body velocities, forward velocity is the dominant component, i.e., $u >> v$ and $u >> w$.
\end{itemize}
Then, using \eqref{eom_final}, equations of motion related to longitudinal dynamics for boost phase become
\begin{equation}
\label{eqn:chapter5_boost_pitch_eqns}
\begin{split}
X &= m\dot{u} + mwq\\
Z &= m\dot{w} - mqu\\
M &= I_y\dot{q}
\end{split}
\end{equation}
Moreover, considering the assumptions above, force and moment terms are simplified as
\begin{equation}
\label{eqn:chapter5_boost_force_eqns}
\begin{split}
X &= QA(C_{x0} + C_{xq}\frac{d}{2u}q)+T-W\sin(\theta)\\
Z &= QA(C_{z0} + C_{zq}\frac{d}{2u}q)-T\theta_T+W\cos(\theta)\\
M &= QAd(C_{m0} + C_{mq}\frac{d}{2u}q)+Tl_x\theta_T
\end{split}
\end{equation}
Substituting \eqref{eqn:chapter5_boost_force_eqns} into \eqref{eqn:chapter5_boost_pitch_eqns}, and considering the additional states related to height, $z$, and pitch angle, $\theta$, boost phase dynamics can be obtained as
\begin{equation}
\label{eqn:chapter5_boost_state_eqns_1}
\begin{split}
\dot{u} & = {(QAC_{x0} + QAC_{xq}\frac{d}{2u}q + T - W\sin(\theta) - mwq)}/{m} \\
\dot{w} & = {(QAC_{z0} + QAC_{zq}\frac{d}{2u}q -T\theta_T+ W\cos(\theta) + mqu)}/{m} \\
\dot{q} & = {(QAdC_{m0} + QAC_{mq}\frac{d^2}{2u}q + Tl_x\theta_T)}/{I_y}\\
\dot{\theta} &= q\\
\dot{z} &= -\sin(\theta)u+\cos(\theta)w
\end{split}
\end{equation}
Then, the longitudinal dynamics of the missile in boost phase can be expressed as
\begin{equation}
\label{eqn:chapter5_boost_state_eqns_final}
\boldsymbol{\dot{x}} = \boldsymbol{h} (\boldsymbol{x},\boldsymbol{u})
\end{equation}
where the state vector $\boldsymbol{x}$ and control vector $\boldsymbol{u}$ are,
\begin{equation}
\begin{split}
\boldsymbol{x} &= [u(t),w(t),q(t),\theta(t),z(t)]^T \\
\boldsymbol{u} &= [T(t), \theta_T(t)]^T
\end{split}
\end{equation}
and $\boldsymbol{h} (\boldsymbol{x},\boldsymbol{u})$ is the vector of expressions given in right hand side of \eqref{eqn:chapter5_boost_state_eqns_1}.

To verify the assumptions made and see how the simplified systems which are given
in \eqref{eqn:chapter5_launch_state_eqns_final} and \eqref{eqn:chapter5_boost_state_eqns_final} approximate to the reference 6 DOF nonlinear dynamics, various scenarios with different control inputs and initial conditions are simulated and the results are compared for nonlinear and simplified models. Through the comparisons it is seen that the simplified models for launch and boost phases are accurate enough relative to the reference nonlinear model and they can be used in optimal control design.

\subsection{Minimum-Effort Optimal Control Problem Formulation}

This class of optimal control problems can be described as of finding an optimal control $\boldsymbol u(t)$ satisfying constraints of the form
\begin{equation}
{M_{i-}} \leq {u_i}(t) \leq {M_{i+}}, \quad i=1,2,...,m
\end{equation}
where ${u_i}(t)$ is the $i^{th}$ control variable, $m$ is the number of control variables, ${M_{i-}}$ and ${M_{i+}}$ are allowed minimum and maximum values of each control variable, and which transfers a system described by
\begin{equation}
\boldsymbol{\dot{x}}(t) = \boldsymbol{h}(\boldsymbol{x}(t),\boldsymbol{u}(t), t)
\end{equation}
from an initial state $\boldsymbol{x}(t_0)$ to a specified final state $\boldsymbol{x}(t_f)$ with a minimum expenditure of control effort \cite{kirk2012optimal}. Control effort to be minimized can be defined by the following performance index:
\begin{equation}
J(\boldsymbol{u}) = \int_{t_0}^{t_f}\bigg[\sum_{i=1}^{m} r_iu{_i}^2(t)\bigg]dt
\end{equation}
where $r_i, i=1,2,...,m$ are nonnegative weighting factors for control variables. For the problem considered in this work, performance index to be minimized is defined as
\begin{equation}
\label{eqn:chapter5_perforamnce_index}
J(\boldsymbol{u}) = \int_{t_0}^{t_f}T^2(t)dt
\end{equation}
so that the square of applied thrust through a time interval, which is a representation of the total energy need, is minimized. To solve the formulated infinite dimensional optimal control problem by numerical methods, it can be transformed into a finite dimensional optimization problem by trapezoidal discretization \cite{speyer2010primer} of the integral in \eqref{eqn:chapter5_perforamnce_index} as
\begin{equation}
\label{eqn:chapter5_cost_discrete}
J(\boldsymbol{u}) = \frac{\Delta t}{2}(T_0^2+2(T_1^2+...+T_{N-1}^2)+T_N^2)
\end{equation}
where $T_k$ (for $k=0,1,2,...,N$) are discrete samples of the applied thrust, and $\Delta t$ is an appropriate time step for discretization. Now, the problem can be considered as a finite dimensional constrained optimization problem where the aim is to find the $m\times(N+1)$ dimensional control input,
\begin{equation}
\boldsymbol{u} = [\boldsymbol{u_0} \boldsymbol{u_1} ... \boldsymbol{u_{N-1}} \boldsymbol{u_N}]
\end{equation}
which minimizes
\begin{equation}
J(\boldsymbol{u}) = \frac{\Delta t}{2}(T_0^2+2(T_1^2+...+T_{N-1}^2)+T_N^2)
\end{equation}
and subject to
\begin{equation}
\begin{split}
&\boldsymbol{x}(t=0)=\boldsymbol{x_0}, \qquad \boldsymbol{x}(t=N\Delta t)=\boldsymbol{x_f}\\
&{\boldsymbol{M_{-}}} \leq \boldsymbol{u} \leq \boldsymbol{M_{+}}
\end{split}
\end{equation}
where ${\boldsymbol{M_{-}}} \in \mathbb{R}^{m\times(N+1)}$ and ${\boldsymbol{M_{+}}} \in \mathbb{R}^{m\times(N+1)}$ are the matrices that define lower and upper limits of the allowed discrete control inputs. 

The derived constrained optimization problem formulation is specialized for launch and boost phases as follows. Since the control variable for the launch phase is the applied thrust, $T(t)$, the constrained optimization problem for this phase can be described as of finding
\begin{equation}
u = [T_0 T_1 ... T_{N-1} T_N]
\end{equation}
which minimizes
\begin{equation}
J({u}) = \frac{\Delta t}{2}(T_0^2+2(T_1^2+...+T_{N-1}^2)+T_N^2)
\end{equation}
subject to
\begin{equation}
\begin{split}
\boldsymbol{x}(t=0) & =[u(0),w(0),q(0),\theta(0),z(0)]^T \\
\boldsymbol{x}(t=t_f) & =[u(t_f),w(t_f),q(t_f),\theta(t_f),z(t_f)]^T\\
0 \leq T_k & \leq 30000 \quad  for \quad k=0,1,2,...,N
\end{split}
\end{equation}

For the boost phase, the control variables are the applied thrust, $T(t)$, and the thrust deflection in the vertical plane, $\theta_T(t)$. Then, the constrained optimization problem for this phase can be described as of finding
\begin{equation}
\boldsymbol{u} = 
\left[
\begin{matrix}
T_0 T_1 ... T_{N-1} T_N \\
{\theta_T}_0 {\theta_T}_1 ... {\theta_T}_{N-1} {\theta_T}_N
\end{matrix}
\right]
\end{equation}
which minimizes
\begin{equation}
J({u}) = \frac{\Delta t}{2}(T_0^2+2(T_1^2+...+T_{N-1}^2)+T_N^2)
\end{equation}
subject to
\begin{equation}
\begin{split}
\boldsymbol{x}(t=0) &=[u(0),w(0),q(0),\theta(0),z(0)]^T \\
\boldsymbol{x}(t=t_f) &=[u(t_f),w(t_f),q(t_f),\theta(t_f),z(t_f)]^T \\
0 \leq &T_k \leq 30000 \quad  for \quad k=0,1,2,...,N \\
-12^{\circ} \leq &{\theta_T}_k  \leq 12^{\circ} \qquad  for  \quad k=0,1,2,...,N
\end{split}
\end{equation}

Available thrust is limited to 30 kN. For the boost phase, the allowed thrust deflection is in the range of [-12,12] degrees. For both of the phases, discretization time step, $\Delta t$, is chosen as 0.2 seconds. This means that for example, in the launch phase, optimal thrust solution to be found for 15 seconds motion, is a vector with 76 elements. These elements can be regarded as parameters to be found by constrained optimization problem solvers. To solve the described constrained optimization problems, MATLAB's Optimization Toolbox is utilized with its $fmincon$ function which is used to find the minimum of constrained nonlinear multivariable functions.

\subsection{Optimal Control Solutions for Launch and Boost Phases}

\subsubsection{Different Water-Exit Pitch Angle Scenarios}

Firstly, different water-exit pitch angle scenarios are investigated. The set of $\theta=\{20, 35, 45, 55, 65, 75, 90\}$ degrees are chosen as the possible water-exit pitch angles.

\begin{figure}[!t]
	\centering
	\subfloat[]{\includegraphics[width=2.6in]{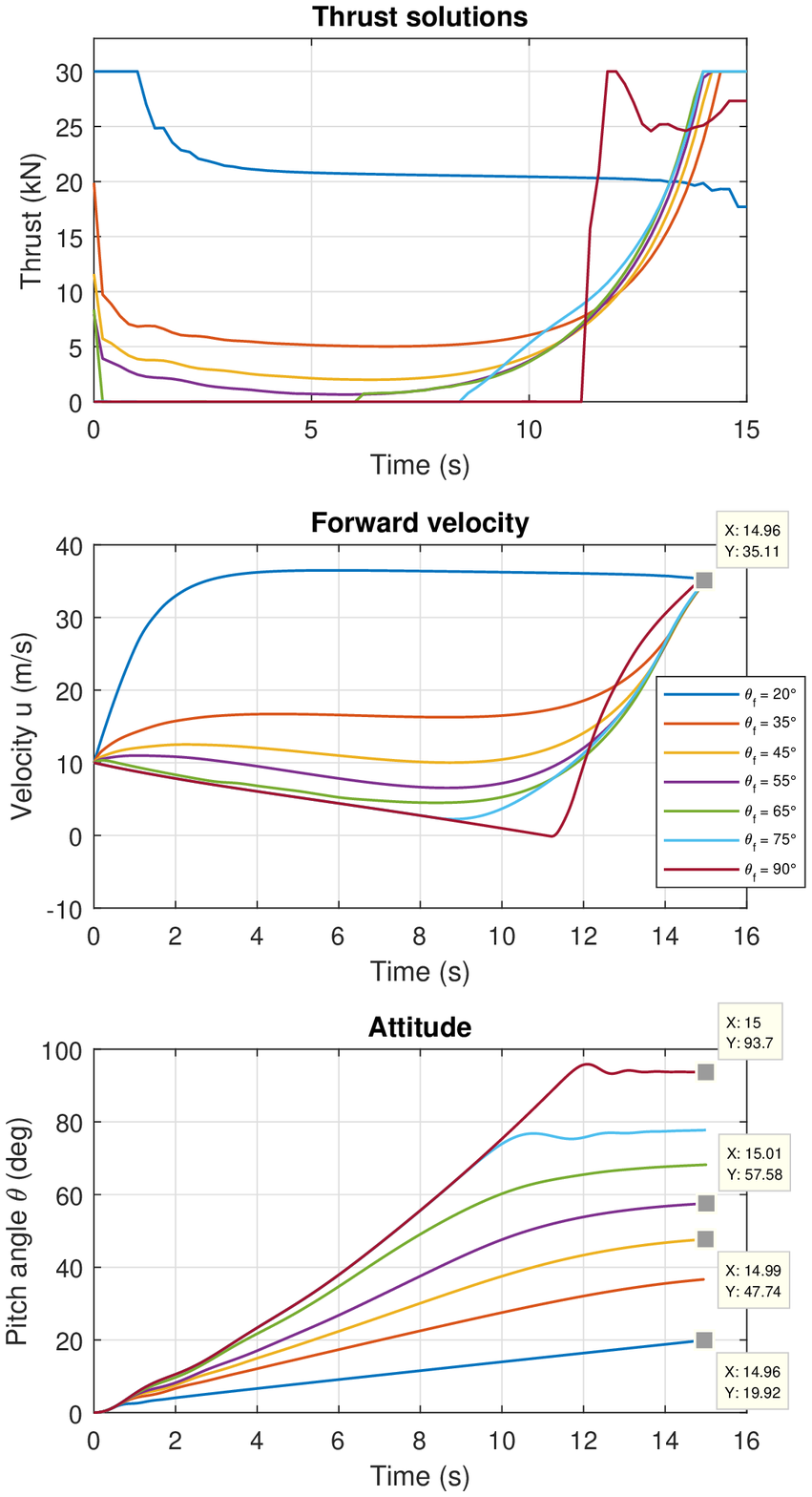}%
		\label{fig_first_case}}
	\hfil
	\subfloat[]{\includegraphics[width=2.6in]{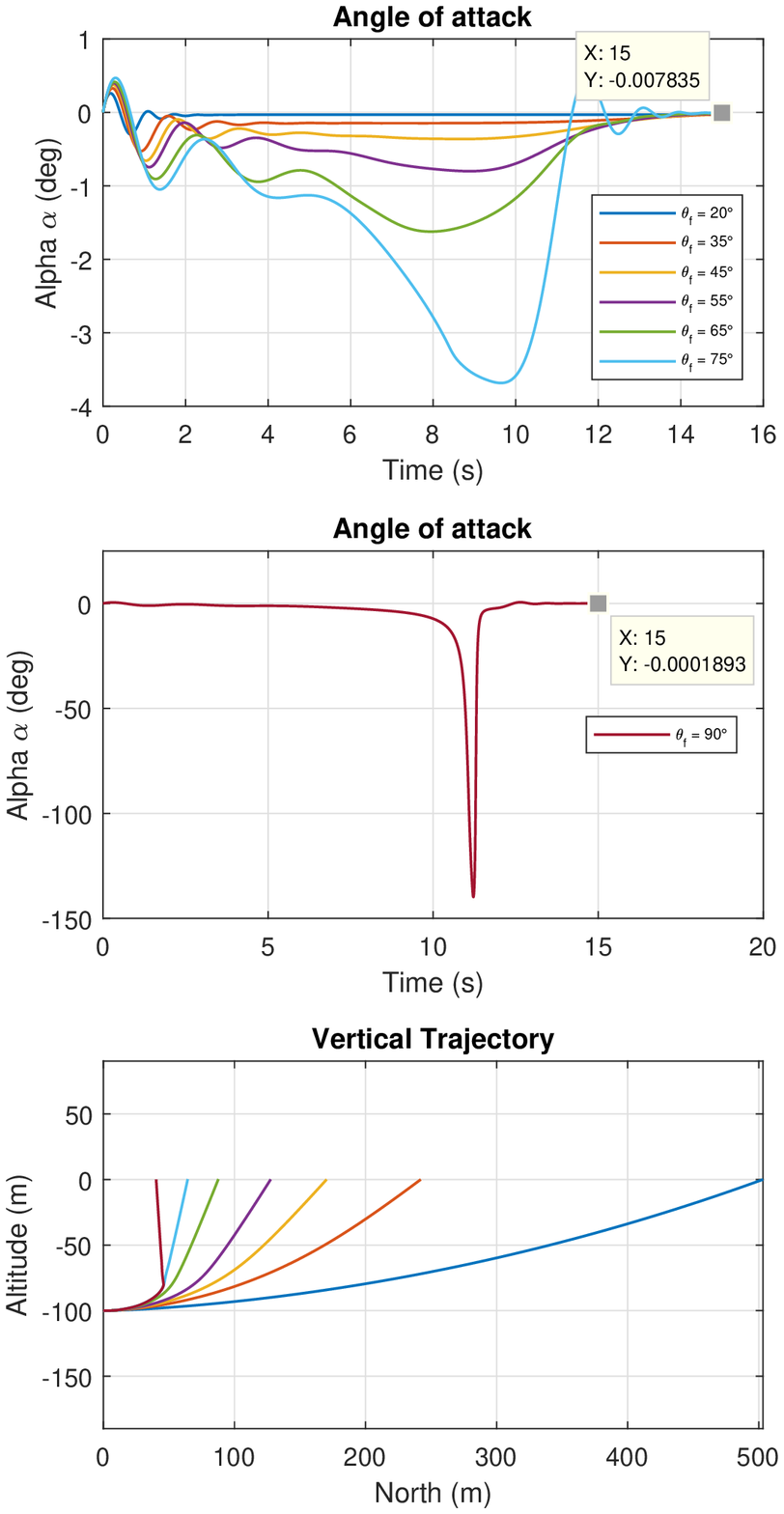}%
		\label{fig_second_case}}
	\caption{Launch phase optimal thrust solutions and the 6 DOF nonlinear simulation results for different water-exit pitch angle scenarios}
	\label{fig:launch_dif_water_exit}
\end{figure}

\underline{Launch Phase Solutions for Horizontal Launch Case:} The conditions just after the launch from the submarine are assigned as the initial conditions of the system at the beginning of this phase. These conditions are described as; 10 m/s forward velocity, 100 m water depth and zero down velocity, pitch rate, and pitch angle. Desired final conditions, which are the water-exit conditions, are assumed as; 35 m/s forward velocity, 0 m water depth and one of the pitch angles from the set of  $\theta=\{20, 35, 45, 55, 65, 75, 90\}$ degrees. Final pitch rate and down velocity are left as free parameters. To choose the duration of the motion, several underwater motion simulations were run by applying different constant thrust inputs for the given initial conditions. Then, the duration of the motion is determined as 15 seconds which is considered to be a suitable motion duration for each water-exit scenario. As a result, seven different scenarios are investigated in this section. Initial and final conditions are taken as
\begin{equation}
\begin{split}
\boldsymbol{x_0} &= [10,0,0,0,100]^T \\
\boldsymbol{x_f} &= [35,free,free,{\theta_f},0]^T\\
\end{split}
\end{equation}
where $\theta_f$ is assigned to the $n^{th}$ element of the set $\{20, 35, 45, 55, 65, 75, 90\}$, for the $n^{th}$ scenario. Optimal thrust solutions and the 6 DOF nonlinear simulation results for each scenario that are obtained by applying the thrust solutions are shown in Fig. \ref{fig:launch_dif_water_exit}. The simulations are stopped when the missile reaches the sea surface. 


Examining the thrust solutions, it is seen that as desired water-exit angle increases, initial thrust levels decrease. Due to the self positive pitching of the missile after the ejection with the given initial conditions, for 75 and 90 degrees solutions, non-zero thrust values are observed around 10 seconds after the ejection. On the other hand, for smaller water-exit angle scenarios, it is needed to apply positive thrust values just after the ejection. When the simulation results are examined, it is seen that in each scenario, the missile reaches the sea surface in almost 15 seconds with the forward velocity of 35 m/s, as desired. As the angle of attack values are zero during water-exit, the total velocity of the missile is equal to the achieved forward velocity, and the total velocity vector is aligned with the positive $x$ axis of the missile body. Water-exit pitch angles are approximately the same as with their desired values. Although there are differences between the obtained and the desired water-exit angles, differences are in acceptable levels. The maximum relative absolute difference is observed in 45-degree-exit scenario. Realized 47.71 degrees water-exit angle corresponds to the relative difference percentage of 6.1\%. This deviation is expected since the simplified launch phase model pitch angle results slightly differ from the nonlinear launch phase model pitch angle results. Vertical trajectory results show that as the desired water-exit angle decreases, the distance covered by the missile through the north increases. This means that to achieve a smaller water-exit angle, a larger sea area around the ejection point is needed.

The cost for each thrust profile is calculated according to \eqref{eqn:chapter5_cost_discrete} and is given in Table \ref{table:chapter5_launch_dif_theta_cost}. The thrust profile that results with minimum cost is found for 45 degrees water-exit angle case.

\begin{table}[!t]
	\renewcommand{\arraystretch}{1.0}
	\caption{Cost of Launch Phase Optimal Thrust Solutions for Different Water-Exit Pitch Angle Scenarios of Horizontal Launch}
	\label{table:chapter5_launch_dif_theta_cost}
	\centering
	\resizebox{.6\textwidth}{!}{%
		\begin{tabular}{||c||c|c|c|c|c|c|c||} 
			\hline
			Scenario no & 1 & 2 & 3 & 4 & 5 & 6 & 7\\
			\hline
			$\theta_f \; (deg)$  & 20 &  35   & 45 & 55 & 65 & 75 & 90\\ 
			\hline
			Cost $(10^9N^2s)$  & 7.03	& 1.80  & 1.67 & 1.72 & 1.78 & 1.82 & 2.45 \\ 
			\hline
		\end{tabular}
	}
\end{table}

\begin{figure}[!t]
	\centering
	\subfloat{\includegraphics[width=2.6in]{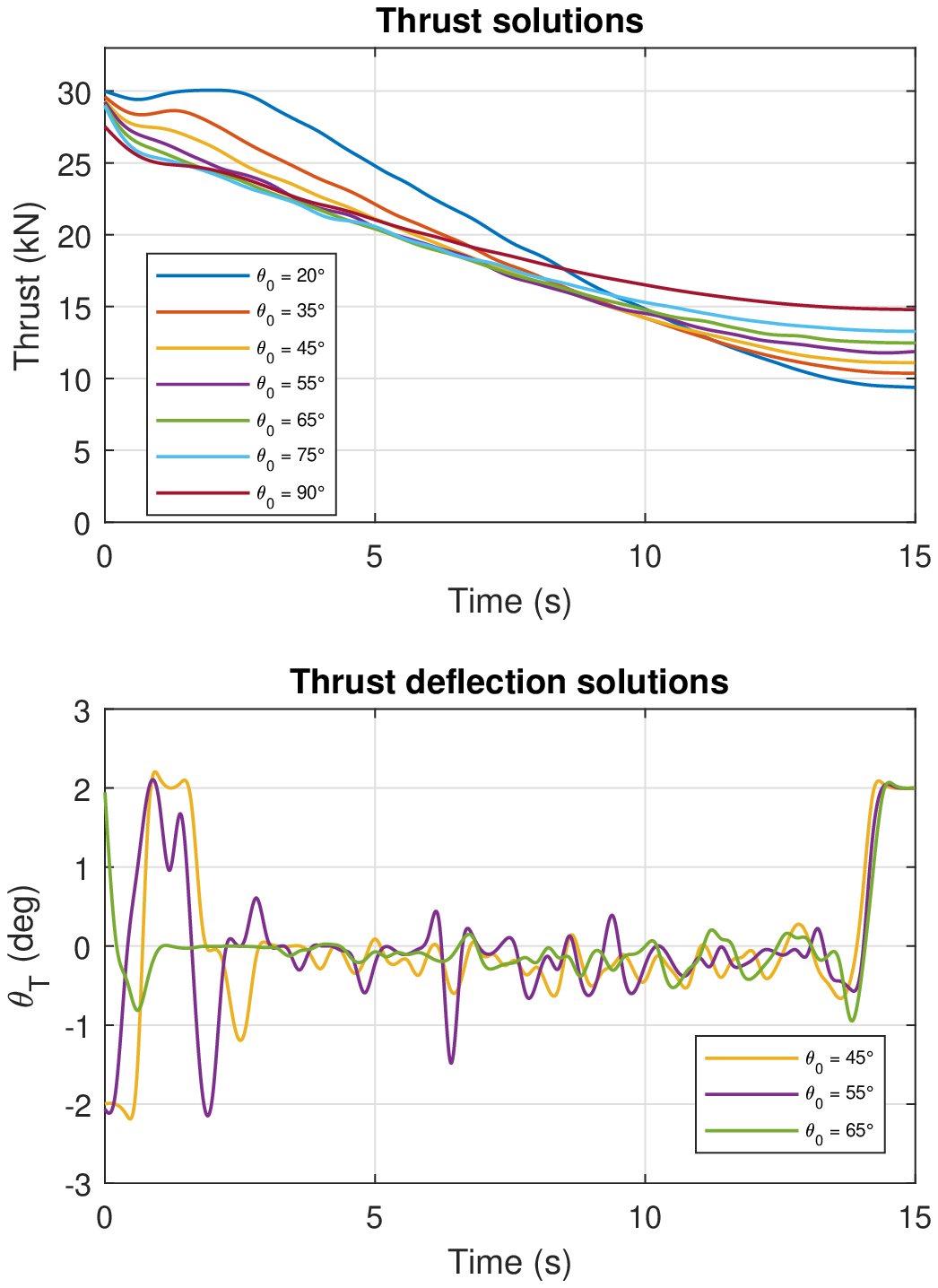}%
	}
	\hfil
	\subfloat{\includegraphics[width=2.6in]{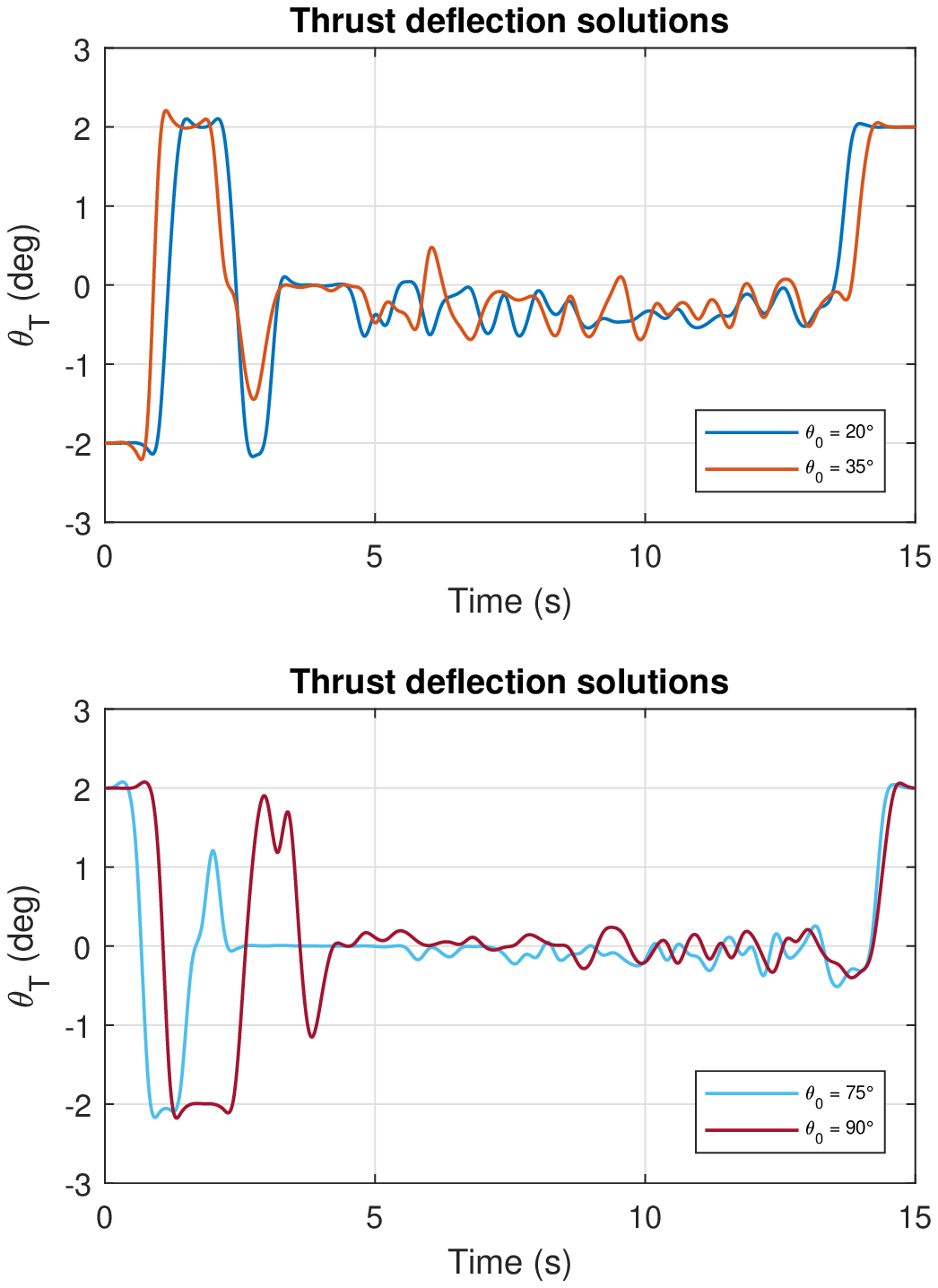}%
	}
	\caption{Boost phase optimal thrust and vertical thrust deflection solutions for different water-exit pitch angle scenarios}
	\label{fig:boost_dif_water_exit_part1}
\end{figure}

\begin{figure}[!t]
	\centering
	\subfloat{\includegraphics[width=2.6in]{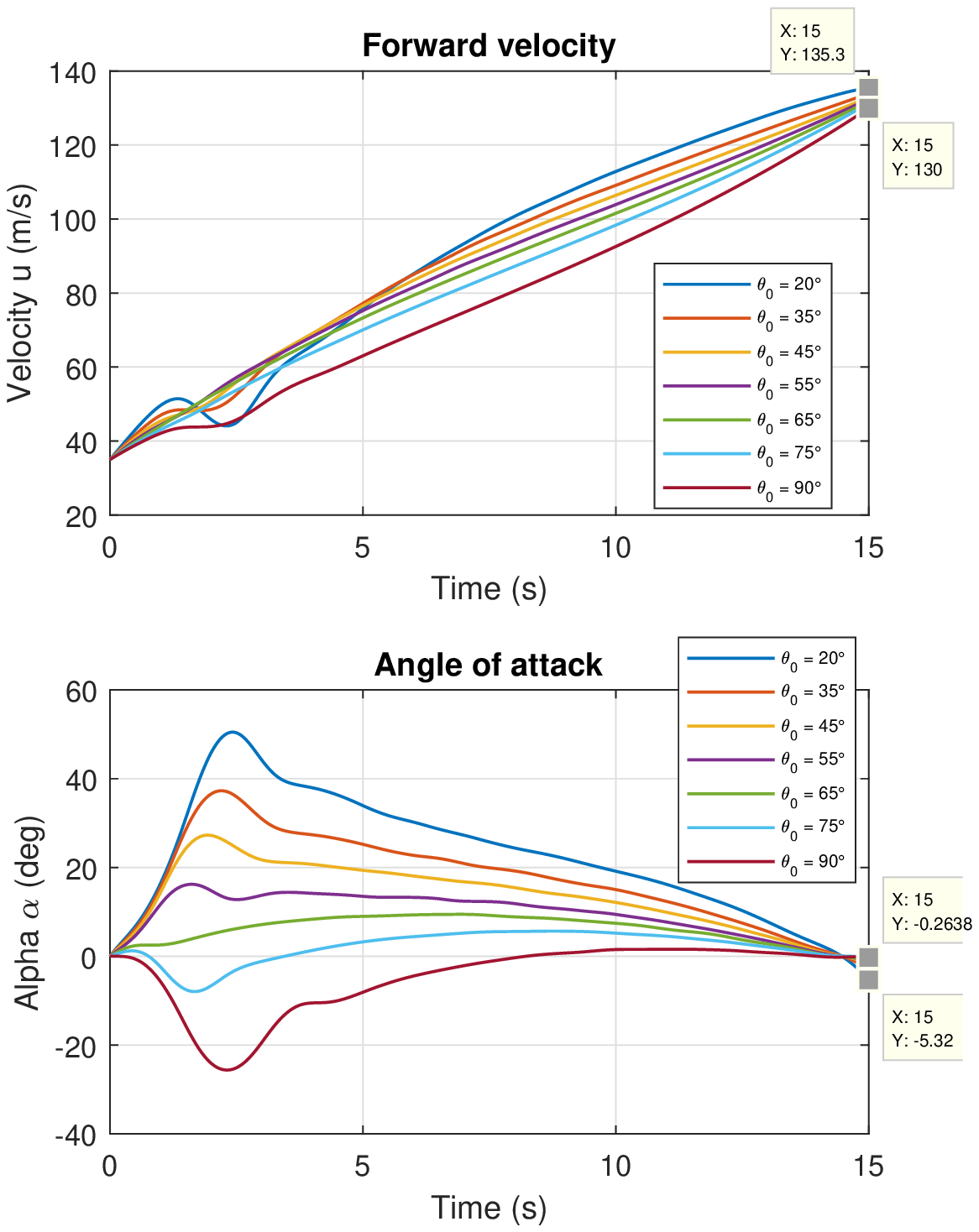}%
	}
	\hfil
	\subfloat{\includegraphics[width=2.6in]{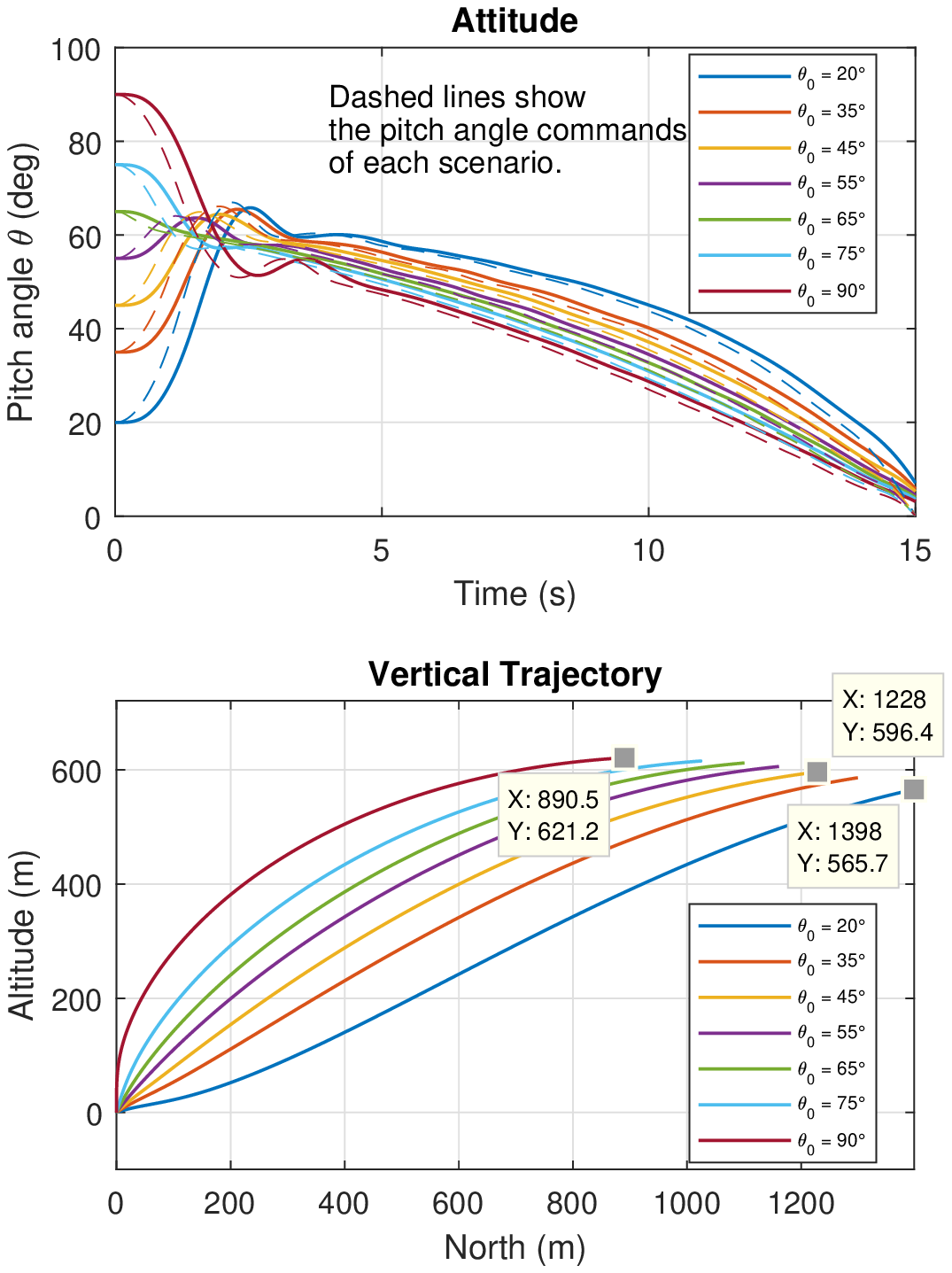}%
	}
	\caption{Boost phase the 6 DOF nonlinear simulation results for different water-exit pitch angle scenarios}
	\label{fig:boost_dif_water_exit_part2}
\end{figure}

\underline{Boost Phase Solutions:} The conditions, just after the missile leaves the sea water, are assigned as the initial conditions of the system at the beginning of the boost phase. Final conditions of the launch phase can be considered as the initial conditions of the boost phase. These desired conditions are: 35 m/s forward velocity and zero altitude, down velocity and pitch rate. Initial pitch angle set, $\theta=\{20, 35, 45, 55, 65, 75, 90\}$, constitutes the seven different scenarios to be investigated here. Desired final conditions are assumed as; 135 m/s forward velocity, zero pitch angle, and 600 m altitude. Final pitch rate and down velocity values are left as free parameters. To determine the motion duration, a similar procedure that is explained for launch phase is followed, and the final time is fixed to 15 seconds. In conclusion, initial and final conditions can be shown as
\begin{equation}
\begin{split}
\boldsymbol{x_0} &= [35,0,0,{\theta_0},0]^T \\
\boldsymbol{x_f} &= [135,free,free,0,600]^T\\
\end{split}
\end{equation}
where $\theta_0$ is assigned to the $n^{th}$ element of the set $\{20, 35, 45, 55, 65, 75, 90\}$, for the $n^{th}$ scenario. Obtained optimal thrust, $T(t)$, and vertical thrust deflection, $\theta_T(t)$, solutions are shown in Fig. \ref{fig:boost_dif_water_exit_part1}. Examining the optimal solutions, it is seen that the thrust solutions have similar characteristics for each case. However, as the water-exit angle increases, the initial thrust levels decrease and the final thrust levels increase. For water-exit angles of 20, 35, 45 and 55 degrees, thrust deflection solutions start with negative values to provide positive pitching moment. On the other hand, for water-exit angles of 65, 75 and 90 degrees, positive starting values of thrust deflection cause negative pitching moment.
Here, to obtain the desired trajectories in the 6 DOF nonlinear simulations, in addition to applying the optimal thrust solutions, pitch angle profiles that are obtained from the simplified model results can be used to guide a pitch angle autopilot. Then, instead of using optimal thrust deflection solutions with an open loop control structure, a closed loop control can be achieved by using simplified model's pitch angle output to feed the pitch angle autopilot of boost phase. Thus, the 6 DOF nonlinear model is simulated with obtained thrust profiles and pitch angle commands for a designed linear quadratic tracker based pitch angle autopilot. Simulations are run for 15 seconds. The simulation results are shown in Fig. \ref{fig:boost_dif_water_exit_part2}.

Comparing the pitch angle results of the nonlinear model and the pitch angle commands, it is seen that final pitch angle result values go nearly to zero, but the exact zero value is not obtained. The reasons for this situation are as follows. First, pitch angle autopilot's command tracking performance is not ideal. Second, the simplified boost phase model pitch angle results slightly differ from the nonlinear boost phase model pitch angle results. However, this deviation can be considered as acceptable by examining how the error in reference pitch angle tracking affects the achievement of other desired final conditions.

Final forward velocity values for different scenarios are in the range of 130 m/s and 135.3 m/s. They are pretty much the same as the desired value of 135 m/s. The maximum relative absolute difference, which is seen in 90 degrees water-exit scenario, is 3.7\%. At the end of the 15 seconds, angle of attack values approximate zero, meaning that the total velocity vector is almost equal to the forward velocity vector. Final altitude values, where the desired final altitude is 600 meters, are in the range of 565.7-621.2 meters, meaning that relative absolute difference percentage is in the range of 3.53-5.72\%, which can be regarded as acceptable. The results of vertical trajectories show that as water-exit angle decreases, the distance covered by the missile through the north increases. This means that after leaving the sea with a smaller water-exit angle, a larger air zone around the water-exit location is needed before the desired final conditions are achieved.

The cost for each scenario is given in Table \ref{table:chapter5_boost_dif_theta_cost}. The thrust profile which has the minimum cost occurs in 65 degrees water-exit angle scenario.

\begin{table}[!t]
	\renewcommand{\arraystretch}{1.0}
	\caption{Cost of Boost Phase Optimal Thrust Solutions for Different Water-Exit Pitch Angle Scenarios}
	\label{table:chapter5_boost_dif_theta_cost}
	\centering
	\resizebox{.6\textwidth}{!}{%
		\begin{tabular}{||c||c|c|c|c|c|c|c||} 
			\hline
			Scenario no & 1 & 2 & 3 & 4 & 5 & 6 & 7\\
			\hline
			$\theta_0 \; (deg)$  & 20 &  35   & 45 & 55 & 65 & 75 & 90\\ 
			\hline
			Cost $(10^9N^2s)$  & 6.75	& 5.82  & 5.46 & 5.30 & 5.27 & 5.35 & 5.76 \\ 
			\hline
		\end{tabular}
	}
\end{table}

\subsubsection{Different Launch Depth Scenarios for Vertical Launch Case} 
\begin{figure}[!t]
	\centering
	\subfloat{\includegraphics[width=2.6in,valign=c]{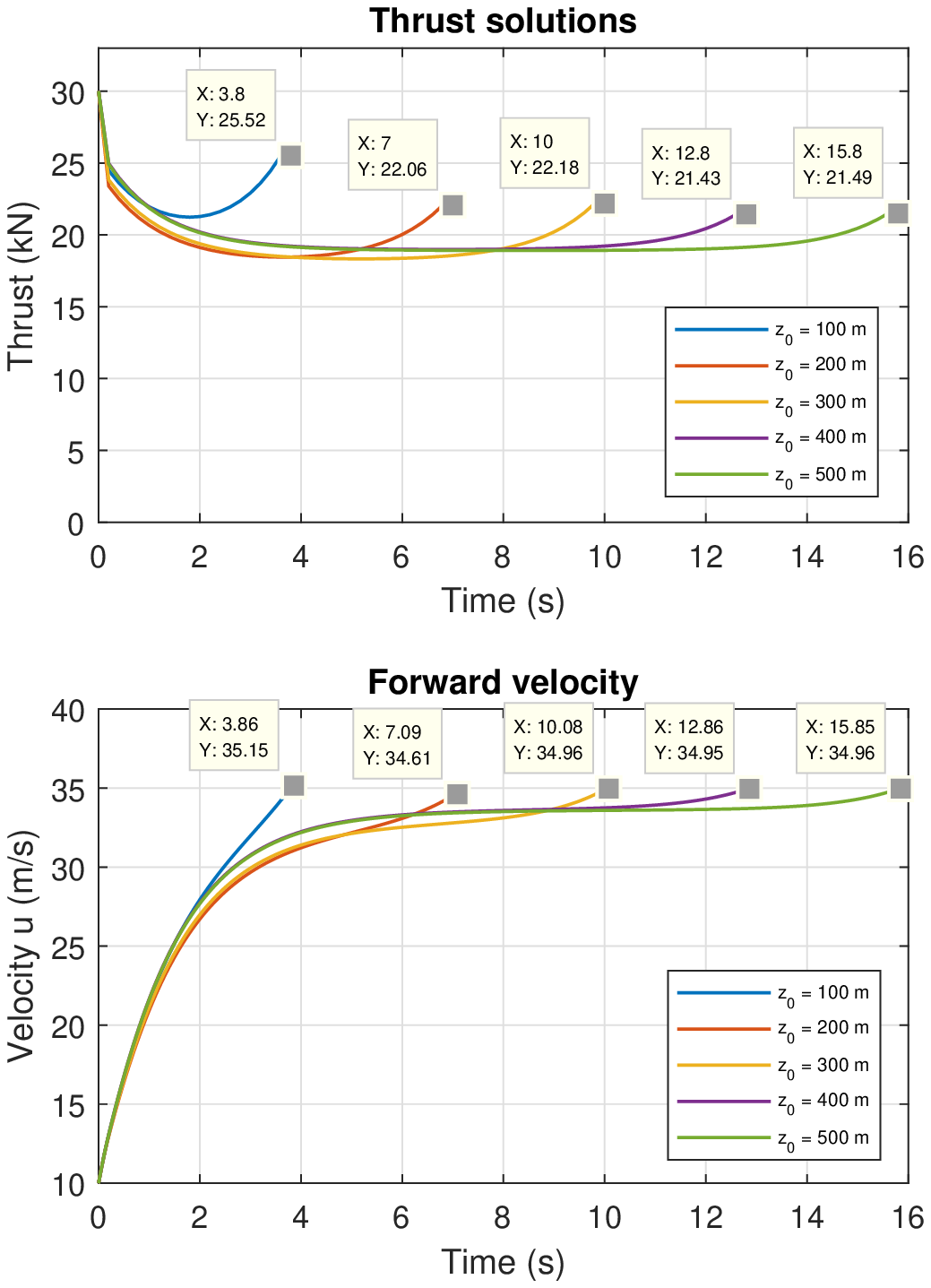}%
	}
	\hfil
	\subfloat{\includegraphics[width=2.6in,valign=c]{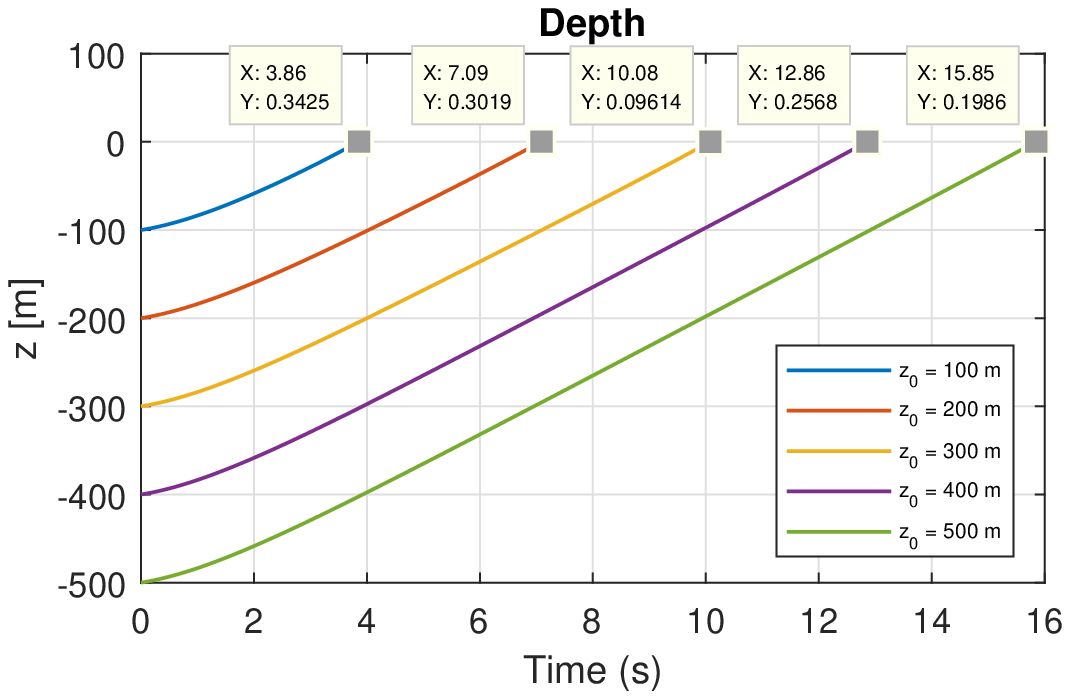}%
	}
	\caption{Launch phase optimal thrust solutions and the 6 DOF nonlinear simulation results for different launch depth scenarios of vertical launch}
	\label{fig:boost_dif_launch_depth}
\end{figure}

For the case of vertical launching from the submarine, it is assumed that the missile leaves the water vertically with a desired forward velocity. Initial conditions are specified as; 10 m/s forward velocity, 90 degrees pitch angle, water depths of $z_0 = \{100, 200, 300, 400, 500\}$ m, zero down velocity and pitch rate. Desired final conditions are assumed as; 35m/s forward velocity, 90 degrees pitch angle, 0 m water depth. Final pitch rate and down velocity are left as free parameters. Thus, five different scenarios are presented. The final time for each scenario is found with the procedure explained as: for each scenario, the magnitude of constant thrust which achieves the desired approximate final conditions is determined. Then, the duration of motion from launch to water-exit is determined to be used in the described five different optimal control scenarios. These final times for five scenarios are $t_f = \{3.8, 7.0, 10.0, 12.8, 15.8\} $ seconds, respectively. In conclusion, initial and final conditions can be shown as;
\begin{equation}
\begin{split}
\boldsymbol{x_0} &= [10,0,0,90,z_0]^T \\
\boldsymbol{x_f} &= [35,free,free,90,0]^T\\
\end{split}
\end{equation}
where $z_0$ is assigned to the $n^{th}$ element of the set $\{100, 200, 300, 400, 500\}$ and $t_f$ is assigned to the $n^{th}$ element of the set $\{3.8, 7, 10, 12.8, 15.8\}$ for the $n^{th}$ scenario. Obtained optimal thrust solutions and the 6 DOF nonliner model simulation results that are achieved by the optimal thrust solutions are shown in Fig. \ref{fig:boost_dif_launch_depth}. 

Since the angle of attack and pitch angle results are constant as 0 and 90 degrees through the simulations, they are not shown in separate figures here. The figure of depth change shows that the motion for each scenario is almost completed in their desired final time as the missile reaches the sea surface. Forward velocity results show that the desired final velocity of 35 m/s is also achieved almost exactly in each scenario.

The cost for each scenario is given in Table \ref{table:chapter5_launch_dif_depth_cost}. As it is expected, the data shows that as the launch depth increases, the cost increases. The thrust profile which provides the minimum cost is obtained for the scenario in which the launch depth is 100 m.

\begin{table}[!t]
	\renewcommand{\arraystretch}{1.0}
	\caption{Cost of Launch Phase Optimal Thrust Solutions for Different Launch Depth Scenarios of Vertical Launch}
	\label{table:chapter5_launch_dif_depth_cost}
	\centering
	\begin{tabular}{||c||c|c|c|c|c||} 
		\hline
		Scenario no & 1 & 2 & 3 & 4 & 5\\
		\hline
		$z_0 \; (m)$  & 100 &  200   & 300 & 400 & 500\\ 
		\hline
		Cost $(10^9N^2s)$  & 1.99	& 2.78  & 3.86 & 5.09 & 6.15\\ 
		\hline
	\end{tabular}
\end{table}

\subsubsection{Cost Analysis for Different Water-Exit Pitch Angle Scenarios of Horizontal Launch and Different Launch Depth Scenarios of Vertical Launch}

In the previous sections, the cost of different water-exit scenarios for launch and boost phases were given. Minimum cost scenarios among them are 45 degrees water-exit scenario of the launch phase and 65 degrees water-exit scenario of the boost phase. Considering the fact that the boost phase follows the launch phase, optimal thrust solutions for launch and boost phases are combined, and the minimum cost calculation is made by using these combined thrust profiles. To accomplish a vertical water-exit mission, the missile can be launched horizontally or vertically from the submarine. Since vertical water-exit scenario of horizontal and vertical launch options can provide different costs, their costs can also be compared to find optimal vertical water-exit mission cost for a set of fixed initial and final conditions. 

\begin{figure}[!t]
	\centering
	\subfloat[]{\includegraphics[width=2.6in,valign=c]{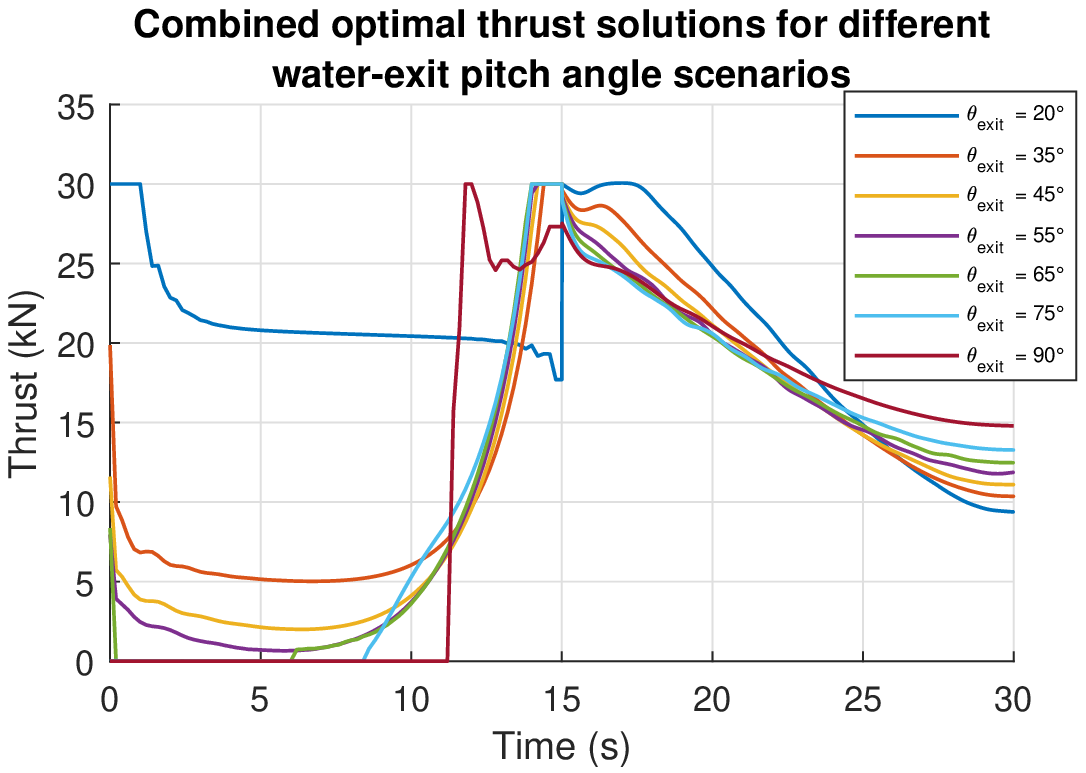}%
	}
	\hfil
	\subfloat[]{\includegraphics[width=2.6in,valign=c]{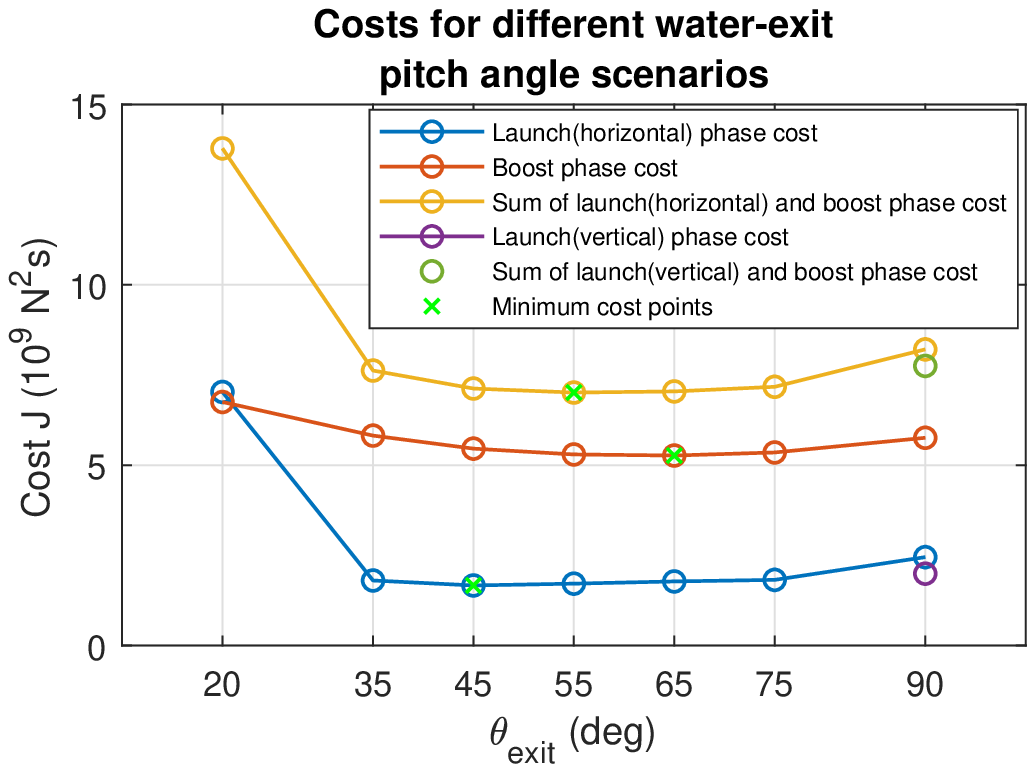}%
	}
	\caption{(a) Launch and boost phase combined optimal thrust solutions for different water-exit pitch angle scenarios (b) The costs of the combined optimal thrust solutions}
	\label{fig:combined_thrust_and_costs_dif_water_exit}
\end{figure}

Combined optimal thrust solutions which are found for horizontal launch phase and boost phase for different water-exit pitch angle scenarios, and the costs of these combined optimal thrust profiles 
are shown in Fig. \ref{fig:combined_thrust_and_costs_dif_water_exit}. In addition to them, the launch phase cost for a vertical launch from 100 m depth, and total energy need as the sum of vertical launch phase and boost phase costs are also shown in the same figure. The data of total costs is given in Table \ref{table:chapter5_cost_analysis_diff_water_exit_angles_cost}. 
\begin{table}[!t]
	\renewcommand{\arraystretch}{1.0}
	\caption{Total Costs of Launch and Boost Phase Optimal Thrust Solutions for Different Water-Exit Pitch Angle Scenarios}
	\label{table:chapter5_cost_analysis_diff_water_exit_angles_cost}
	\centering
	\resizebox{.6\textwidth}{!}{%
		\begin{tabular}{||c||c|c|c|c|c|c|c||} 
			\hline
			Scenario no & 1 & 2 & 3 & 4 & 5 & 6 & 7\\
			\hline
			$\theta_{exit} (deg)$  & 20 &  35   & 45 & 55 & 65 & 75 & 90\\ 
			\hline
			Total launch(horizontal) and boost   & &   & &  &  & &\\
			phase costs $(10^9N^2s)$   & 13.78	& 7.62  & 7.13 &7.01 &7.05 & 7.17 & 8.21  \\ 
			\hline
			Total launch(vertical) and boost  & 	&   &  &  &  &  &  \\ 
			phase costs $(10^9N^2s)$ & - & - & - & -  & -  & - & 7.75 \\ 
			\hline
		\end{tabular}
	}
\end{table}

According to the data of total costs, the following conclusions can be made. For the previously defined initial and final conditions of launch and boost phases, 55 degrees water-exit angle scenario yields the minimum total cost, which is $7.01\times10^9 N^2s$. This cost is also less than the total cost of a vertical launch and boost phase, which is $7.75\times10^9 N^2s$. For the vertical water-exit case, total cost of the vertical launch and boost phase, which is $7.75\times10^9 N^2s$, is less than the total cost of the horizontal launch and boost phase, which is $8.21\times10^9 N^2s$. Thus, for the given initial and final conditions of launch and boost phases, the following deductions can be made:

\begin{itemize}
	\item Among the examined water-exit angles, horizontal launch scenario aiming 55 degrees water-exit angle yields minimum energy need.
	
	\item If vertical water-exit is desired, vertical launch from the submarine needs less energy than that of horizontal launch from the submarine.
	
\end{itemize}

\subsubsection{Effects of Other Initial and Final Conditions on Cost}

The effect of desired water-exit angle on total cost and flight results are investigated in previous parts. In these scenarios, initial and final conditions of forward velocity, time and depth/altitude are fixed. In this part, it is examined how the variations of these conditions affect the cost. For the launch phase, variations in final forward velocity, initial depth and final time are investigated. For the boost phase, variations in final forward velocity, final altitude and final time are investigated. For each condition, variation sets are determined as the values for which there are feasible solutions.

\begin{figure}[!t]
	\centering
	\subfloat[]{\includegraphics[width=2.6in,valign=c]{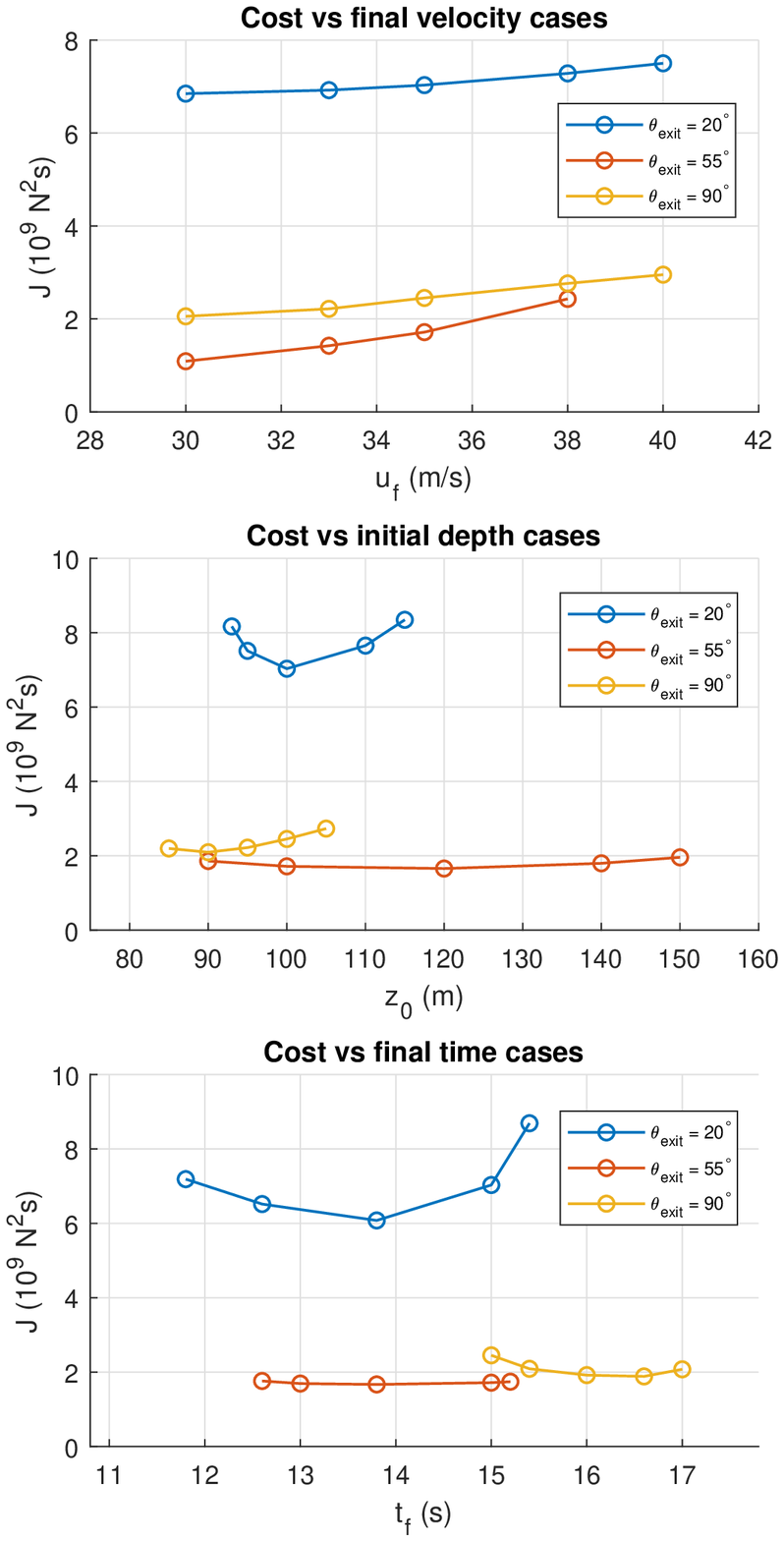}%
		\label{fig:cond_vars_launch}
	}
	\hfil
	\subfloat[]{\includegraphics[width=2.6in,valign=c]{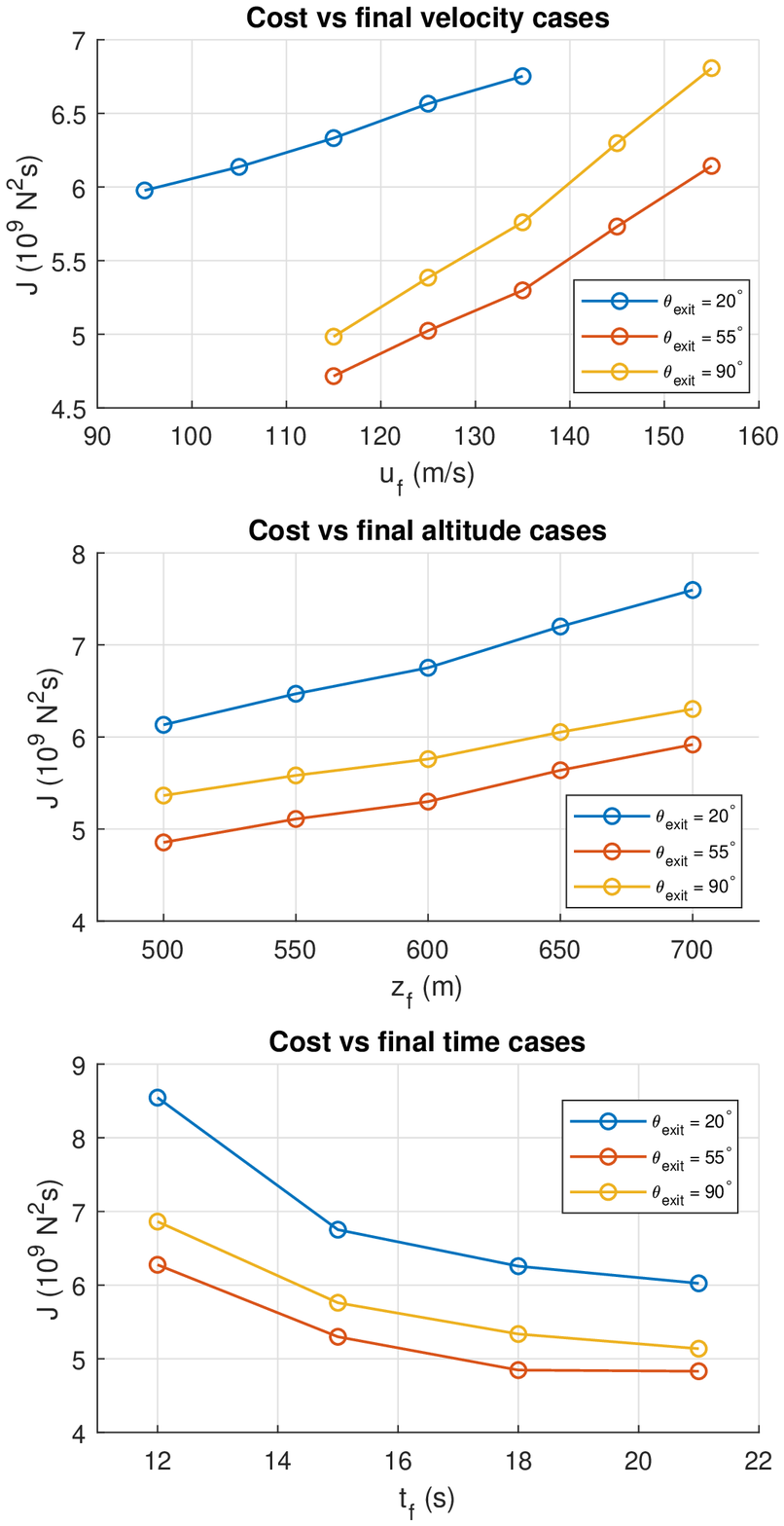}%
		\label{fig:cond_vars_boost}
	}
	\caption{(a) Launch phase cost change with final velocity, initial depth and final time variations for $\theta_{exit} = \{20, 55, 90\}$ cases (b) Boost phase cost change with final velocity, final altitude and final time variations for $\theta_{exit} = \{20, 55, 90\}$ cases}
	\label{fig:cond_vars_launch_and_boost}
\end{figure}

For all of the cases, cost results for water-exit scenarios of $\theta_{exit}=\{20,55,90\}$ are considered. For the launch phase, initial condition is $\boldsymbol{x_0} = [10,0,0,0,100]^T$, except for the case where initial depth variations are examined. Final condition is ${x_f} = [35,free,free,\theta_{exit},0]^T$, except for the case where final forward velocity variations are examined. Final time is fixed to 15 seconds, except for the case where final time variations are examined. For the boost phase, initial condition is $\boldsymbol{x_0} = [35,0,0,\theta_{exit},0]^T$. Final condition is $\boldsymbol{x_f} = [135,free,free,0,600]^T$ for the case where final time variations are examined. Final time is fixed to 15 seconds for the cases where final forward velocity and altitude are examined. In Fig. \ref{fig:cond_vars_launch_and_boost} it is shown that how the cost changes with the variations in  initial and final conditions. Results can be summarized as:
\begin{itemize}
	\item  For the launch phase, as final forward velocity increases, the total energy need increases. Interestingly, however, it is not necessarily true that the cost increases or decreases as initial depth or final time increases, but there are optimal initial depth and final time values which promise minimum energy need for the examined cases. 
	\item For the boost phase, as final forward velocity and final altitude increase, the total energy need increases. As final time increases, the cost decreases. 
\end{itemize}

\subsubsection{Determination of Optimal Initial and Final Conditions for Launch and Boost Phases}
So far, it is investigated how the initial and final conditions of the launch and boost phases affect the total cost. Using some evidence of that investigation as a starting point, these initial and final conditions can be optimized to obtain a mission motion profile which needs minimum energy. In the following sections; firstly, optimal initial depth, final forward velocity, and final time conditions are found for the launch phase of horizontal launch. Secondly, optimal final altitude, final forward velocity, and final time conditions are found for the boost phase. These optimal values are found for the water-exit angle set of $\{45, 55, 65\}$ degrees. Then, analyzing the horizontal launch phase and boost phase optimal costs for different water-exit angles together, the optimal water-exit angle is determined. Besides that, optimal initial depth, final forward velocity, and final time conditions are determined for the launch phase of vertical launch. \newline

\underline{Optimal Conditions for Launch Phase:}
\begin{itemize}
	\item {Horizontal Launch Case}
\end{itemize}
To find optimal final forward velocity value, first the initial conditions are fixed to standard horizontal launch phase conditions, which is $\boldsymbol{x_0} = [10,0,0,0,100]^T$. As different from the previous cases, final forward velocity is considered as a parameter to be found by the optimal control solver. Instead of searching only discrete samples of applied thrust, the optimization algorithm is implemented to search both applied thrust and final forward velocity. Other final conditions are determined as; pitch angle set of $\{45, 55, 65\}$ degrees, zero altitude and free choice of pitch rate and down velocity. It is seen, independent of the final pitch angle, this method gives the optimal final forward velocity as its allowed minimum value provided to the optimal control solver. This result is consistent with the observation which states that as final forward velocity increases, the total energy need increases for the launch phase. So, the optimal final forward velocity value can be selected according to another limiting factor related to the system or the mission. In this work, it is fixed to 35 m/s.

\begin{figure}[!t]
	\centering
	\subfloat[]{\includegraphics[width=2.6in,valign=c]{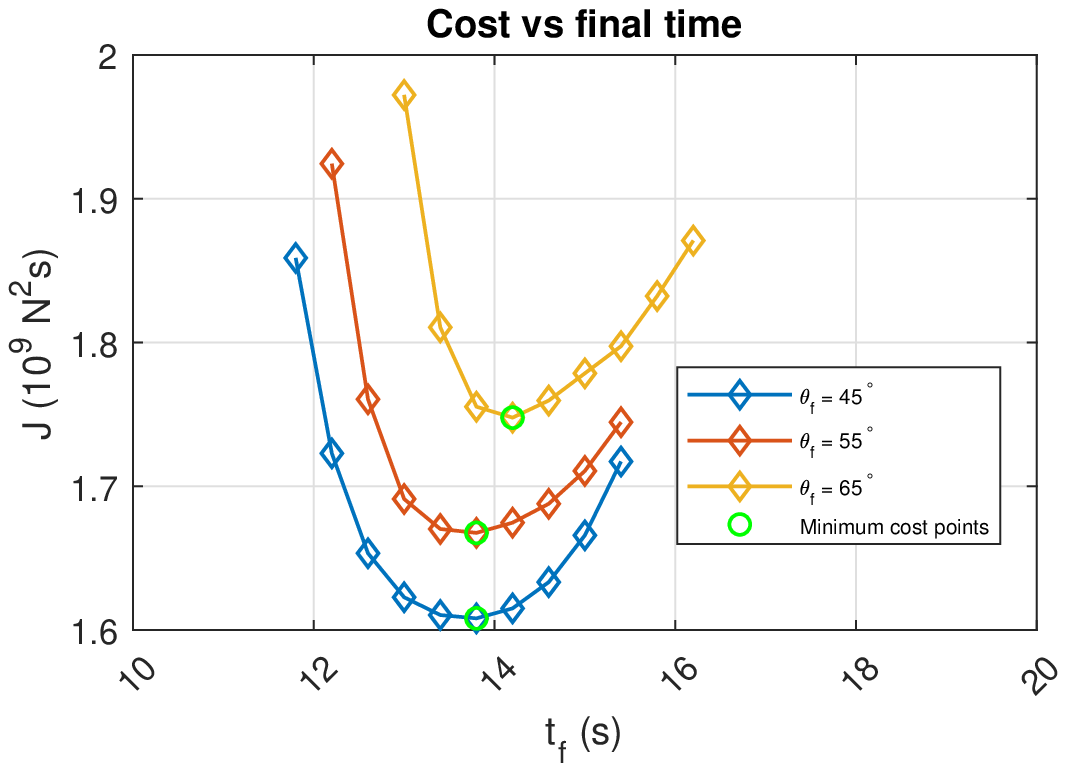}%
		\label{fig:opt_conds_launch_cost_vs_tf_a}
	}
	\hfil
	\subfloat[]{\includegraphics[width=2.6in,valign=c]{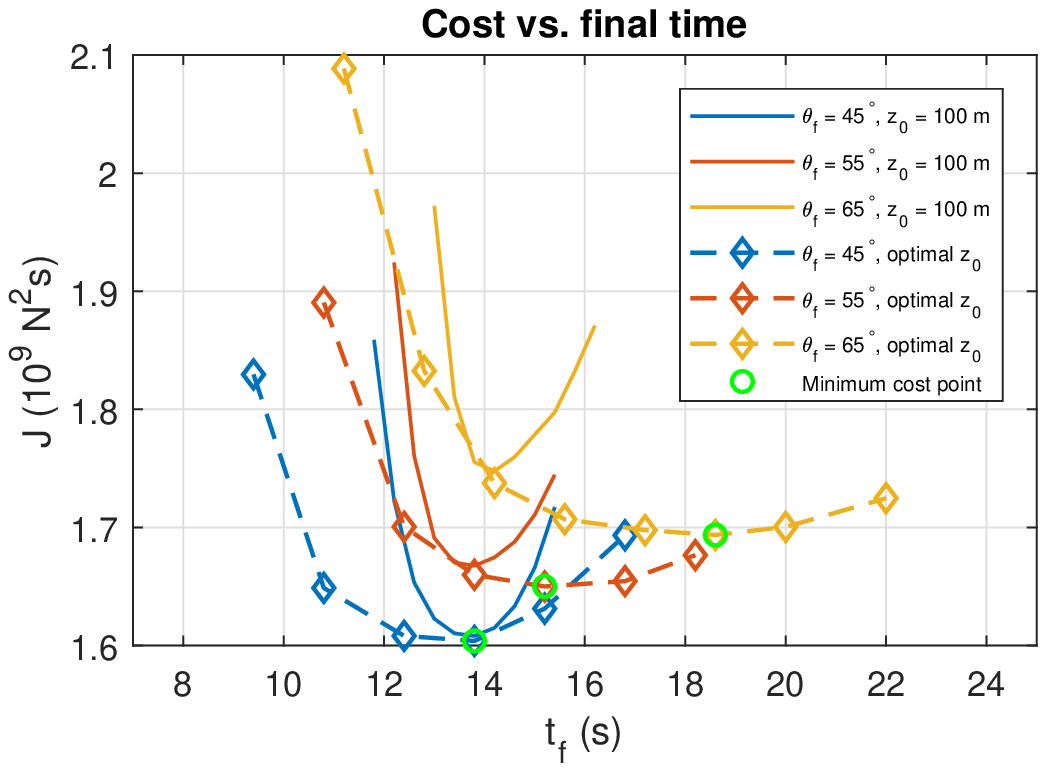}%
		\label{fig:opt_conds_launch_cost_vs_tf_b}
	}
	\caption{(a) Launch phase cost change with final time, for fixed initial depth, for $\theta_{f} = \{45, 55, 65\}$ cases (b) Launch phase cost change with final time, for fixed and optimal initial depth,  for $\theta_{f} = \{45, 55, 65\}$ cases}
	\label{fig:opt_conds_launch_cost_vs_tf}
\end{figure}

Different final time conditions are investigated to find optimal final time candidates. Initial and final conditions are fixed to $\boldsymbol{x_0} = [10,0,0,0,100]^T$ and $\boldsymbol{x_f} = [35,free,free,\theta_f,0]^T$, respectively. $\theta_f$ is an element of the set $\{45, 55, 65\}$. Cost change with final time for each water-exit pitch angle scenario are shown in Fig. \ref{fig:opt_conds_launch_cost_vs_tf_a}. 
The final times which provide minimum cost are 13.8, 13.8 and 14.2 seconds for 45, 55 and 65 degrees water-exit pitch angle scenarios, respectively.

%
%

Optimal initial depth values are searched for the previously found final time values, which provide minimum cost, and in the neighborhood of them. The procedure followed to find optimal final forward velocity is also used here, but this time, optimal control solver is implemented to search for applied thrust and initial depth which minimizes the performance index. Initial condition is fixed to $\boldsymbol{x_0} = [10,0,0,0,z_0]^T$ where $z_0$ is to be found by the optimal control solver. Final conditions are fixed to $\boldsymbol{x_f} = [35,free,free,\theta_f,0]^T$ where $\theta_f$ is an element of the set $\{45, 55, 65\}$. In Fig. \ref{fig:opt_conds_launch_cost_vs_tf_b}, cost change with final time for each water-exit pitch angle scenario with fixed initial depth and optimal initial depth are given. The cost data for different water-exit pitch angles with optimal initial depth values are also given in Table \ref{table:chapter5_opt_conds_launch_cost_vs_tf_with_opt_z0}. From the table, it is seen that for given final conditions, the final time and initial depth values which provide minimum cost are, 13.8 seconds and 101.78 meters, 15.2 seconds and 118.75 meters, 18.6 seconds and 149.29 meters for 45, 55 and 65 degrees water-exit pitch angle scenarios, respectively. At this point it should be noticed that optimal final time candidates are also updated while searching for the optimal initial depths.

Among three different water-exit pitch angle scenarios whose cost is optimal with respect to the initial depth, final forward velocity and final time, 45 degrees pitch angle scenario provides minimum cost. However, to decide which water-exit pitch angle should be chosen as optimal for complete mission profile, it must be analyzed together with boost phase results of the next section.

\begin{table}[!t]
	\renewcommand{\arraystretch}{1.0}
	\caption{Launch Phase Costs for Different Final Time, for Optimal Initial Depth, for $\theta_{f} = \{45, 55, 65\}$ Cases}
	\label{table:chapter5_opt_conds_launch_cost_vs_tf_with_opt_z0}
	\centering
	\resizebox{.6\textwidth}{!}{%
		\begin{tabular}{||c||c|c|c|c|c|c|c|c||} 
			\hline
			$ \theta_f(deg) $ & \multicolumn{8}{c||}{45} \\
			\hline
			$ t_f(s) $ & 9.4 & 10.8 & 12.4 & 13.8 & 15.2 & 16.8 & - & - \\
			\hline
			$ z_0(m) $ & 55.41 & 70.60 & 88.93 & 101.78 & 112.12 & 128.22 & - & - \\
			\hline
			$ J(10^9N^2s) $ & 1.830 & 1.649 & 1.608 & 1.604 & 1.631 & 1.693 & - & - \\
			\hline \hline
			
			$ \theta_f(deg) $ & \multicolumn{8}{c||}{55} \\
			\hline
			$ t_f(s) $ & 10.8 & 12.4 & 13.8 & 15.2 & 16.8 & 18.2 & - & - \\
			\hline
			$ z_0(m) $ & 67.05 & 88.42 & 105.38 & 118.75 & 131.63 & 146.15 & - & - \\
			\hline
			$ J(10^9N^2s) $ & 1.890 & 1.701 & 1.660 & 1.650 & 1.655 & 1.677 & - & - \\
			\hline \hline
			
			$ \theta_f(deg) $ & \multicolumn{8}{c||}{65} \\
			\hline
			$ t_f(s) $ & 11.2 & 12.8 & 14.2 & 15.6 & 17.2 & 18.6 & 20.0 & 22.0 \\
			\hline
			$ z_0(m) $ & 60.83 & 85.84 & 104.46 & 121.037 & 138.77 & 149.29 & 163.26 & 185.57 \\
			\hline
			$ J(10^9N^2s) $ & 2.089 & 1.833 & 1.738 & 1.707 & 1.698 & 1.694 & 1.700 & 1.725 \\
			\hline
	\end{tabular}}
\end{table}

\begin{itemize}
	\item {Vertical Launch Case}
\end{itemize}

\begin{figure}[!t]
	\centering
	\includegraphics[width=2.6in]{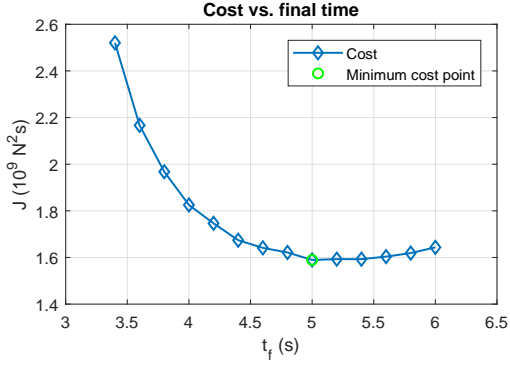}
	\caption{Launch (vertical) phase cost change with final time}
	\label{fig:opt_conds_launch_cost_vs_tf_vertical}
\end{figure}

When final forward velocity and initial depth values are considered as parameters to be found by the optimal control solver, it is seen that the final forward velocity and initial depth values which provide minimum cost are found as their allowed minimum values provided to the optimal control solver. So, final forward velocity and initial depth values are fixed to 35 m/s and 100 meters. To find the final time value which results with the minimum cost, different final time scenarios are analyzed for initial condition $\boldsymbol{x_0} = [10,0,0,90,100]^T$ and final condition $\boldsymbol{x_f} = [35,0,0,90,0]^T$. Cost change with final time for each scenario is shown in Fig. \ref{fig:opt_conds_launch_cost_vs_tf_vertical}. The optimal thrust profile with minimum energy need is the one that causes 5 seconds of underwater motion.
\newline

\underline{Optimal Conditions for Boost Phase:}\\ In order to find optimal final forward velocity and final altitude values, they are considered as parameters to be found by the optimal control solver. The solver is implemented such that it finds discrete samples of applied thrust, thrust deflection, final forward velocity and final altitude values which minimize the performance index. Initial conditions are defined as $\boldsymbol{x_0} = [35,0,0,\theta_0,0]^T$  where $\theta_0$ is an element of the set $\{45, 55, 65\}$. Final pitch angle condition is fixed to zero, where final pitch rate and down velocity are left as free parameters. In the results,
regardless of the initial pitch angle, optimal final forward velocity and final altitude values are found as their allowed minimum values provided to the optimal control solver. This result is consistent with the observation made during the investigation of the effects of the initial and final conditions on cost, which states that for boost phase, as final forward velocity and final altitude increase, the total energy need increases. So, optimal final forward velocity and final altitude values can be selected according to another limiting factor related to the system or the mission. An example of this limiting factor is that, at the beginning of the cruise phase, the missile may need to be within a certain altitude-speed envelope to start the cruise engine. In this work, final forward velocity and final altitude conditions of boost phase are fixed to 135 m/s and 600 m.

The effect of variations in final time on cost is analyzed previously, for the initial condition $\boldsymbol{x_0} = [35,0,0,\theta_0,0]^T$, where $\theta_0$ is an element of the set $\{20, 55, 90\}$, and for the final condition $\boldsymbol{x_f} = [135,0,0,0,600]^T$. It is observed that as final time increases in an interval such that there is still a feasible optimal control solution, total energy need decreases. So, upper limit of the final time for $\theta_0 = \{45,55,65\}$ cases at which there is a feasible solution, can be determined as the optimal final time values. These values are found as 18, 21 and 21 seconds for 45, 55 and 65 degrees water-exit pitch angle cases, respectively. Obtained minimum cost data for different water-exit pitch angles and final time are given in Table \ref{table:opt_conds_boost_cost_vs_theta}.

Among the three different water-exit pitch angle scenarios whose cost is optimal with respect to the final forward velocity, final altitude and final time, 65 degrees pitch angle scenario provides minimum cost. However, to decide which water-exit pitch angle should be chosen as optimal for complete mission profile, in the next section, it is analyzed together with the launch phase results.
\begin{table}[!t]
	\renewcommand{\arraystretch}{1.0}
	\caption{Boost Phase Minimum Costs for $\theta_{0} = \{45, 55, 65\}$ Cases}
	\label{table:opt_conds_boost_cost_vs_theta}
	\centering
	\resizebox{.4\textwidth}{!}{%
		\begin{tabular}{||c||c|c|c||} 
			\hline
			$ \theta_0(deg) $ &45 & 55 & 65  \\
			\hline
			$ t_f(s) $ & 18 & 21 & 21 \\
			\hline
			$ J(10^9N^2s) $ & 5.1551 & 4.8299 & 4.8228 \\
			\hline
		\end{tabular}
	}
\end{table}

\subsection{Cost Analysis for Overall Design}

\begin{figure}[!t]
	\centering
	\subfloat[]{\includegraphics[width=2.6in,valign=c]{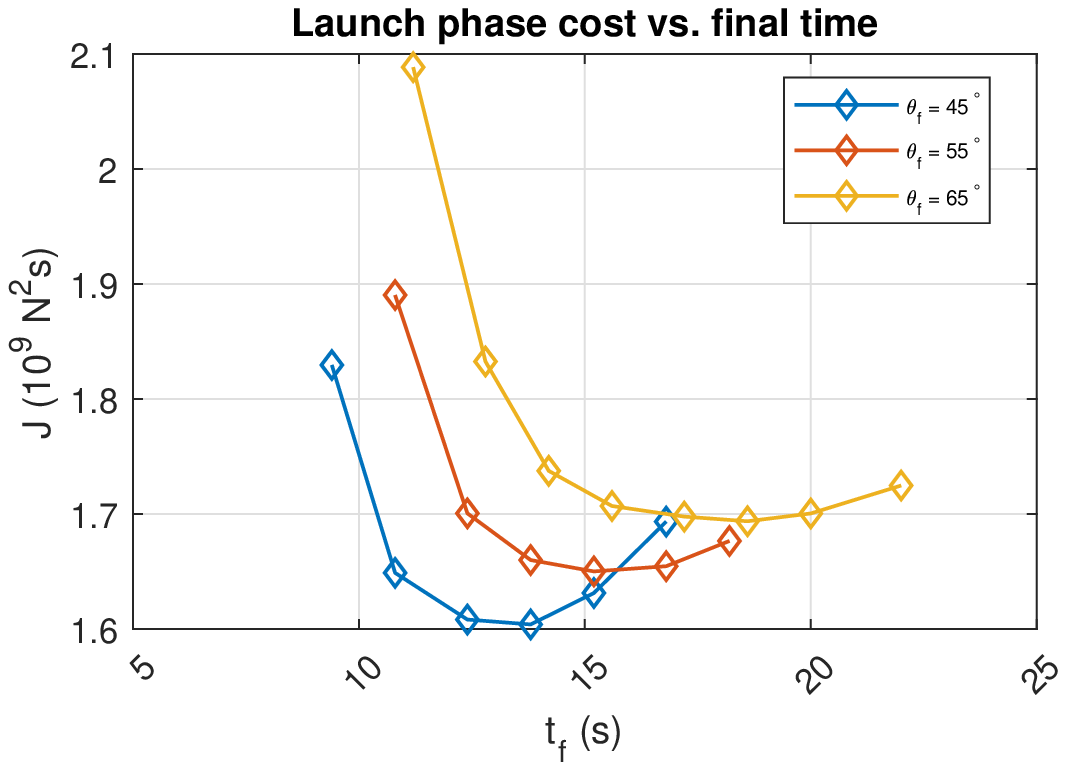}%
		\label{fig:opt_conds_launch_cost_vs_tf_launch_and_total_a}
	}
	\hfil
	\subfloat[]{\includegraphics[width=2.6in,valign=c]{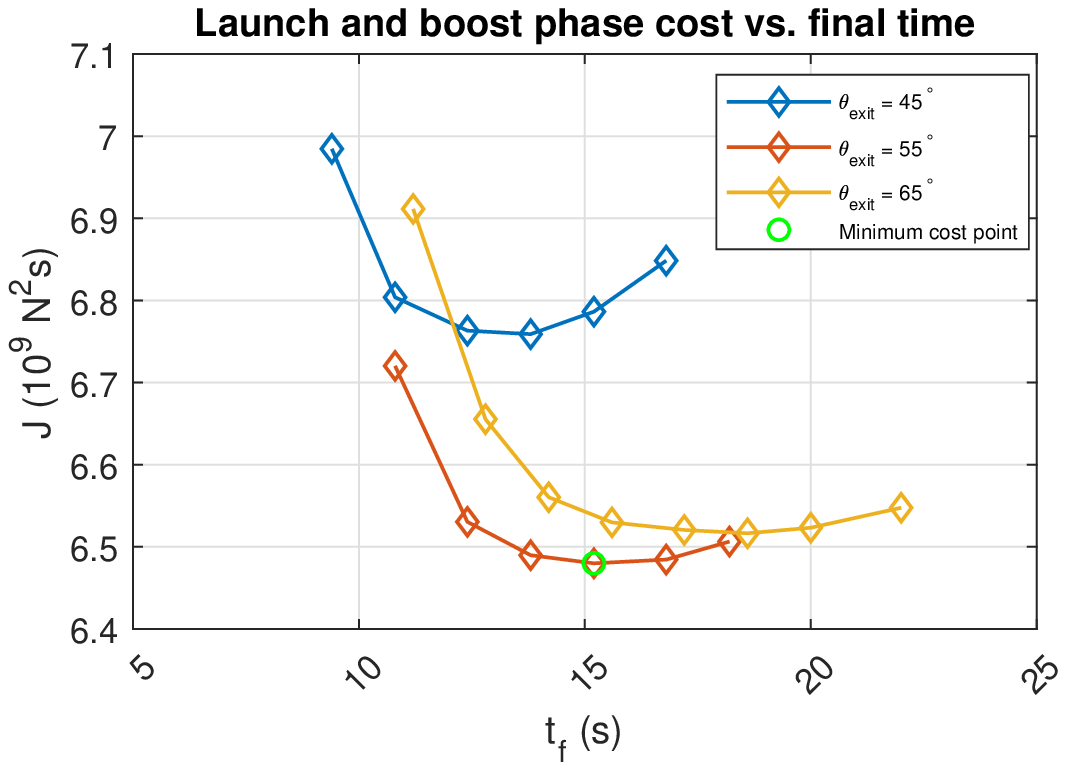}%
		\label{fig:opt_conds_launch_cost_vs_tf_launch_and_total_b}
	}
	\caption{(a) Launch phase cost change with final time, for optimal initial depth, optimal final forward velocity, and for $\theta_{f} = \{45, 55, 65\}$ cases (b) Launch and boost phase cost change with final time, for optimal initial depth, optimal final forward velocity, and for $\theta_{f} = \{45, 55, 65\}$ cases}
	\label{fig:opt_conds_launch_cost_vs_tf_launch_and_total}
\end{figure}

\begin{figure}[!t]
	\centering
	\includegraphics[width=2.6in]{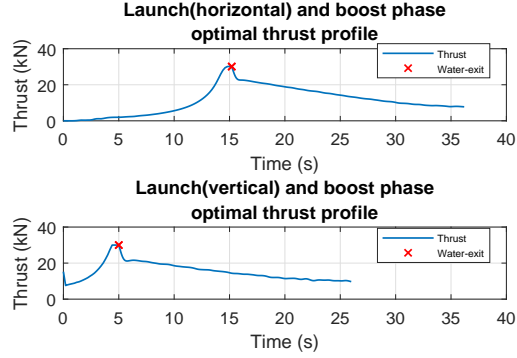}
	\caption{Launch (horizontal and vertical) and boost phase optimal thrust solutions}
	\label{fig:opt_conds_thrust_profiles}
\end{figure}

So far, the optimal inital and final conditions for the launch and boost phases are determined for different water-exit pitch angle scenarios. To decide the optimal water-exit pitch angle condition, the minimum costs obtained for water-exit angles of $\{45, 55, 65\}$ degrees for both launch and boost phases are analyzed together in this part. Fig. \ref{fig:opt_conds_launch_cost_vs_tf_launch_and_total_a} shows launch phase cost change with final time for optimal initial depth and optimal final forward velocity. Fig. \ref{fig:opt_conds_launch_cost_vs_tf_launch_and_total_b} shows the sum of boost phase minimum costs that are given in Table \ref{table:opt_conds_boost_cost_vs_theta}  with the costs in left hand side. So, it is concluded that when launch phase minimum costs and boost phase minimum costs for water-exit pitch angles of $\{45, 55, 65\}$ degrees are added to each other, the minimum cost occurs at 55 degrees water-exit pitch angle. 
Optimal thrust profiles which accomplish the minimum energy scenarios for horizontal launch and vertical launch, to be described below, are given in Fig. \ref{fig:opt_conds_thrust_profiles}. The costs related to the given thrust profiles and corresponding initial and final conditions and completion times are shown in Table \ref{table:chapter5_overall_cost_analysis}. According to the results given in this table, the mission profile for horizontal and vertical launch which need minimum total energy can be described as follows: For a horizontal launch starting with 10 m/s forward velocity, zero pitch angle, pitch rate and down velocity, the missile is launched at a depth of 118.75 m to reach the sea surface with 35 m/s forward velocity and 55 degrees pitch angle in 15.2 seconds. Then, boost phase ends with 135 m/s forward velocity, zero pitch angle at 600 m altitude, in 21 seconds. For a vertical launch starting with 10 m/s forward velocity, zero pitch rate and down velocity at a depth of 100 m, the missile reaches the sea surface with 35 m/s forward velocity in 5 seconds. Then, boost phase ends with 135 m/s forward velocity, zero pitch angle at 600 m altitude, in 21 seconds.

\begin{table}[!t]
	\renewcommand{\arraystretch}{1.0}
	\caption{Optimal Conditions and Costs for Minimum-Energy Horizontal and Vertical Launch Scenarios}
	\label{table:chapter5_overall_cost_analysis}
	\centering
	\resizebox{.8\textwidth}{!}{%
		\begin{tabular}{||c|c|c||} 	
			\hline
			& Horizontal launch case & Vertical launch case \\
			\hline
			Launch phase initial condition, $\boldsymbol{x_0}$ & $[10,0,0,0,118.75]^T$ & $[10,0,0,90,100]^T$ \\
			\hline
			Launch phase final condition, $\boldsymbol{x_f}$  &$[35,free,free,55,0]^T$ & $[35,free,free,90,0]^T$ \\
			\hline
			Launch phase completion time (s) & 15.2 & 5.0 \\
			\hline
			Boost phase initial condition, $\boldsymbol{x_0}$ & $[35,0,0,55,0]^T$ & $[35,0,0,90,0]^T$ \\
			\hline
			Boost phase final condition, $\boldsymbol{x_f}$ & $[135,free,free,0,600]^T$ & $[135,free,free,0,600]^T$ \\
			\hline
			Boost phase completion time (s) & 21.0 & 21.0 \\
			\hline
			Launch phase cost($10^9N^2s$) & 1.6500 & 1.5895 \\
			\hline
			Boost phase cost($10^9N^2s$) & 4.8299 & 5.1358 \\
			\hline
			Total cost($10^9N^2s$) & 6.4799 & 6.7253 \\
			\hline
	\end{tabular}}
\end{table}

\section{Conclusion}
\label{sec:s5}
In this work, energy-optimal guidance and control problem of a submarine-launched cruise missile is studied. The guidance and control of the launch and boost phases are examined as a minimum-effort optimal control problem where the aim is to complete these phases for given initial and final conditions while minimizing the energy need, which is represented as a function of the applied thrust in this study. In addition to finding optimal control solutions that satisfy the given control objectives, the effect of initial and final conditions of the launch and boost phases on energy need is investigated, and optimal conditions which minimize the energy need are determined. This investigation is performed among the set of feasible solutions. 

For the launch phase, which covers the motion from submarine-launch to water-exit, effect of launch depth, final forward velocity, water-exit angle, and final time is investigated. For the horizontal launch scenarios, it is observed that, as the final forward velocity increases, the energy need increases. However, for the other conditions, the energy need does not monotonically increase or decrease, and optimal conditions should be searched in the region of interest. For the vertical launch scenarios, as the launch depth increases the energy need increases as it is expected. Considering only the launch phase of horizontal launch, for the fixed final forward velocity of 35 m/s, optimal conditions are found as 45 degrees water-exit angle, 101.78 m launch depth and 13.8 seconds completion time. Considering the launch phase of vertical launch, for the fixed final forward velocity of 35 m/s, initial depth of 100 m, and vertical water-exit, optimal completion time is found as 5 seconds. For the boost phase, which covers the motion from water-exit to the beginning of the cruise, the effect of final altitude, final forward velocity, water-exit angle, and final time is investigated. It is seen that, as the final forward velocity and final altitude increase, the energy need increases, and as the final time increases, the energy need decreases. However, for the water-exit angle, the energy need does not monotonically increase or decrease, and the optimal value should be searched in the region of interest. Considering only the boost phase, for the fixed final forward velocity of 135 m/s, and the final altitude of 600 m, optimal water-exit angle is found as 65 degrees and optimal completion time is found as 21 seconds. Considering the total energy need of the launch and boost phases together by summing them, the water-exit angle which provides the minimum energy need is found as 55 degrees. Then, for the launch phase, for the fixed final forward velocity of 35 m/s and 55 degrees water-exit angle, 118.75 m launch depth and 15.2 seconds completion time are found as the optimal values. For the boost phase, for the fixed final forward velocity of 135 m/s, final altitude of 600 m and 55 degrees water-exit angle, 21 seconds completion time is found as the optimal value. In the results, it is also seen that the optimal horizontal launch strategy needs less energy than that of the optimal vertical launch strategy. Results show that, whereas some optimal initial and final conditions can be determined according to the system and/or mission constraints, others can be chosen utilizing the minimum-effort optimal control solutions.

A possible improvement which can be focused on in the future works could be related to the booster thrust model. In this study, it is assumed that the desired booster thrust is ideally realized within an upper and lower bound. However, in practice, there are some limitations about the total available thrust, some possible delays during the thrust generation, some characteristic thrust increase or decrease profiles or possible misalignments. These models can be combined with a fuel consumption, dynamically changing mass, inertia, and center of gravity models. Then, a more realistic design can be achieved.

\end{document}